\documentclass[journal]{IEEEtran}

\usepackage{float}
\usepackage{algorithm}
\usepackage{placeins}
\usepackage{algpseudocode}
\usepackage{graphicx}
\usepackage{epstopdf}
\usepackage{scalefnt}
\usepackage[noadjust]{cite}
\usepackage[table]{xcolor}
\usepackage[cmex10]{amsmath}
\usepackage{etoolbox}
\usepackage[framemethod=TikZ]{mdframed}
\usepackage{amssymb}
\usepackage{caption}
\usepackage[labelformat=simple]{subcaption}
\usepackage[bottom]{footmisc}

\newcommand{\Q}{{\vphantom{-1}}}
\newcommand*\rc{\color[rgb]{0, 0, 0}}

\newcounter{subeqn}

\bstctlcite{IEEEexample:BSTcontrol}

\IEEEoverridecommandlockouts
\begin{document}
\title{$M$-QAM Precoder Design for \\ MIMO Directional Modulation Transceivers}
\author{
	\IEEEauthorblockN{Ashkan Kalantari, Christos Tsinos, Mojtaba Soltanalian, \\ Symeon Chatzinotas, Wing-Kin Ma, Erik G. Larsson, and Bj\"{o}rn Ottersten}
	\thanks{\scriptsize This work was supported by the National Research Fund (FNR) of Luxembourg under FNR CORE ECLECTIC and FNR INTER CI-PHY. Ashkan Kalantari and Erik G. Larsson are with Dept. of Electrical Engineering, Link\"{o}ping University, Sweden, ({E-mails:\{ashkan.kalantari,erik.g.larsson\}@liu.se}). Christos Tsinos, Symeon Chatzinotas, and Bj\"{o}rn Ottersten are with SnT, University of Luxembourg, Luxembourg, ({E-mails:\{christos.tsinos,symeon.chatzinotas,bjorn.ottersten\}@uni.lu}). Mojtaba Soltanalian is with the Dept. of Electrical and Computer Engineering, University of Illinois at Chicago, Chicago, (E-mail:msol@uic.edu). Wing-Kin Ma is with the Dept. of Electronic Engineering, The Chinese University of Hong Kong, Shatin, Hong Kong, China, (e-mail:wkma@ieee.org).}}	
\maketitle
%
%
%
\begin{abstract} 
Spectrally efficient multi-antenna wireless communication systems are a key challenge as service demands continue to increase. At the same time, powering up radio access networks is facing environmental and regulation limitations. In order to achieve more power efficiency, we design a directional modulation precoder by considering an $M$-QAM constellation, particularly with $M=4,8,16,32$. First, extended detection regions are defined for desired constellations using analytical geometry. Then, constellation points are placed in the optimal positions of these regions while the minimum Euclidean distance to adjacent constellation points and detection region boundaries is kept as in the conventional $M$-QAM modulation. For further power efficiency and symbol error rate similar to that of fixed design in high SNR, relaxed detection regions are modeled for inner points of $M=16,32$ constellations. The modeled extended and relaxed detection regions as well as the modulation characteristics are utilized to formulate symbol-level precoder design problems for directional modulation to minimize the transmission power while preserving the minimum required SNR at the destination. In addition, the extended and relaxed detection regions are used for precoder design to minimize the output of each power amplifier. We transform the design problems into convex ones and devise an interior point path-following iterative algorithm to solve the mentioned problems and provide details on finding the initial values of the parameters and the starting point. Results show that compared to the benchmark schemes, the proposed method performs better in terms of power and peak power reduction as well as symbol error rate reduction for a wide range of SNRs.
\end{abstract}
\begin{keywords}
Directional modulation, extended detection region, power efficiency, spatial peak power, $M$-QAM modulation, symbol-level precoding. 
\end{keywords}
\section{Introduction}   \label{sec:intro}
Cisco predicts that the mobile data traffic will increase eleven-fold from $2015$ to $2020$~\cite{cisco}. The most important elements which have contributed to this surge in mobile traffic are the extreme growth of video content on the Internet, the advent of mobile devices, e.g., smart phones and tablets, and the market appetite for them. Conventional approaches such as orthogonal frequency division multiplexing access and time division multiplexing access~\cite{OFDMA:1971,TDMA:1995} are used to utilize the frequency and time resources to improve the data communication rate. Later, Multiple-input Multiple-output (MIMO) communication systems emerged, e.g., Long Term Evolution (LTE), to use the spatial domain. MIMO systems provide spatial degrees of freedom in the design at the expense of interference among the transmitted data streams. Pre/post-processing at the transmitter and/or receiver ends are employed to reduce the interference among the data streams~\cite{Spencer:2004,Sidiropoulos:2006}. Recently, there has been a growing interest in directional modulation~\cite{Babakhani:2008,Daly:2009,Daly:2010,DM:JSTSP:kalantari:2016,Kalantari:psk:relax,Kalantari:MQAM,lowcom:pre} and symbol-level precoding for constructive interference~\cite{Masouros:2009,CDMA:Mas:2010,cons:2015,con:glob:2015,multi:adaptive:TWC,pre:survey} to mitigate interference in MIMO systems. In directional modulation, the channel realization and the symbols are used to design the antenna weights. These weights are designed such that the Radio Frequency (RF) signals get modulated after passing through the channel and result in communicating the desired symbol at the desired direction (antenna). {\rc In other words, the symbols are not sent by the transmitter, rather the symbols are induced at the corresponding receiving antennas}. Depending on the design, this can result in no interference~\cite{DM:JSTSP:kalantari:2016,Kalantari:MQAM} or limited interference~\cite{Daly:2009,Kalantari:psk:relax} among the communicated data streams. Both digital symbol-level precoding and directional modulation focus on multiplexing gain. Directional modulation and digital symbol-level precoding for constructive interference differ in the following way. The former focuses on applying array weights in the analog domain to have the desired amplitude and phase for the received signals with design degrees of freedom equivalent to number of antennas. On the other hand, the latter uses symbol-level precoding for digital signal design at the transmitter to create constructive interference among the transmitted data streams at the receiver with design degrees of freedom equivalent to the product of transmitting and receiving antennas.

Apart from an increasing data demand, wireless communications consume a large amount of power and have a considerable share in environmental pollution~\cite{mobile:carbon:2011}. Not only reducing the power consumed in the radio access networks is environmental friendly, but it also decreases the communication costs for both the operators and users~\cite{mob:ene:2010}. The research works in~\cite{Masouros:2009,CDMA:Mas:2010,Masouros:2015,DM:JSTSP:kalantari:2016,EE:const:2016} study the relaxed design in constructive interference and directional modulation with the goal of reducing the power consumption at the transmitter.

Although directional modulation offers transmission with no or limited interference as well as power efficiency, the hardware limitations at the transmitter need to be considered in the precoder design process. Among the hardware limitations, we focus on keeping each power amplifier outside its saturation region to avoid nonlinear distortion of the amplified signal~\cite{amplifier:1981}. 
To do so, we need to design the antenna weights such that the output of each power amplifier avoids the saturation region. In this direction, the references~\cite{Erik:2012:CE,CE:Ma:2014} consider constant envelope precoding for a single-user massive MIMO system. The constant envelop design is restrictive but results in keeping the power amplifier in the desired operating point. The authors in~\cite{danilo:glob:2016,slp:nonlin} consider a low peak power to average ratio design based on constructive interference for $M$-PSK modulation to limit the amplifier output power below a specific value where there is a strict constraint on the phase of the received symbols. A relaxed low peak power design for $M$-PSK modulation in order to include both power efficiency and hardware limitation is proposed in~\cite{Kalantari:psk:relax}.
\subsection{Contributions and Main Results}
Based on the above descriptions, we tackle the design of a system which jointly takes into account the user demand, power amplifier linearity, and {\rc power} efficiency when a finite-alphabet input is considered. The designed precoders for Gaussian input signals can be used to precode finite-alphabet inputs, however, this may result in considerable system performance reduction~\cite{lin:pre:TWC:2012}.

The recent works of~\cite{Masouros:2009,CDMA:Mas:2010,Masouros:2015,EE:const:2016,DM:JSTSP:kalantari:2016,Kalantari:psk:relax} focus on $M$-PSK precoder design. However, there is no work on designing the directional modulation precoder for $M$-QAM constellation, while jointly utilizing the extended and relaxed detection regions of the constellation as well as controlling the power of the amplifier output signal to avoid saturation. To pursue this design, we use the concept of directional modulation and bring the following contributions:
\begin{enumerate}
	\item We define the extended detection regions of $M$-QAM modulation and model these regions using analytical geometry. By extended detection region, we mean a region in which a constellation point can be placed, given that it preserves the standard Euclidean distance with the adjacent constellation points and the detection boundaries. Here, the Euclidean distance among the constellation points in conventional $M$-QAM is considered as the standard distance. The works~\cite{con:glob:2015,multi:adaptive:TWC} consider symbol-level precoder design for $M=16$.
	
	\item In addition to extended detection regions, we characterize relaxed detection regions where the inner points of constellations, here for $M=16,32$, can move within a square-shaped region. Using the extended detection region, we formulate a trade-off between power efficiency at the transmitter and symbol error rate (SER) at the receiver. Particularly, the relaxed design results in SER that is very close to fixed design in high SNR but with less power consumption, specifically for $M=32$. The research~\cite{Kalantari:MQAM} performs directional modulation precoder design for $M$-QAM by only using extended detection region without considering peak power minimization. The works~\cite{con:glob:2015,multi:adaptive:TWC} design the precoder by considering fixed positions for the inner points of $M=16$ constellation. 
	
	\item We define the optimal $M$-QAM directional modulation precoder design problems using the characterized extended detection regions for $M$-QAM to minimize the transmission power while satisfying the required SNR at the antennas of the receiver. Furthermore, we re-design the optimal precoder while considering the relaxed detection regions for inner constellation points and investigate the power efficiency and SER.  
	
	\item We define the optimal precoder design problem to minimize the instantaneous peak power of the amplifier output signal for each RF chain of the directional modulation transmitter. We refer to this as spatial peak power minimization defined as
	\begin{align}
	P_{\max }^{spatial} = \mathop {\max }\limits_{k = 1,...,{N_t}} \,\,{{\bf{w}}^H}{{\bf{E}}_k}{\bf{w}}.
	\label{eqn:SPP}
	\end{align}
	This design is carried out using the characterized extended detection regions for $M$-QAM while preserving the required SNR at the antennas of the receiver. Also, we repeat the precoder design while considering both relaxed and extended detection regions. In~\cite{danilo:glob:2016}, the spatial peak power minimization is carried out for $M$-PSK modulation without considering extended and relaxed detection regions. 
	
	\item The design problems are transformed into convex ones and we devise a fully-detailed interior point path-following iterative algorithm to solve all formulated design problems. In addition, we provide details on finding the starting point and initial values of optimization parameters.    
	
	\item Through extensive simulations, we reveal the benefits of using extended detection regions for $M$-QAM directional modulation transmission in terms of power consumption reduction at the transmitter and SER at the receiver. As a hardware limitation, we evaluate the performance in terms of the mentioned metrics when minimizing the instantaneous spatial peak power. In addition, we quantify the trade-off between power consumption and SER for relaxed detection regions and show their benefit in high SNR while these are not investigated in~\cite{con:glob:2015,multi:adaptive:TWC}.   
\end{enumerate}

The design without relaxed regions translates into interference-free communication since the symbols keep the standard Euclidean distance. 
On the other hand, the precoder design when considering relaxed inner points for $M$-QAM translates into communication with interference.

It is worth mentioning that the precoder design for $128$-QAM is similar to $32$-QAM. Furthermore, the precoder design for $64$-QAM and $256$-QAM are similar to $16$-QAM. Hence, we consider precoding design for $M=4,8,16,32$ constellation points in this work due to lack of space. From a practical point of view, $M$-QAM modulations up to $64$-QAM are used in the long term evolution (LTE) standard~\cite{LTE}.
\subsection{Paper Organization}
The remainder of this paper is organized as follows. In Section~\ref{sec:sys:mod}, we introduce the signal and system model. The extended and relaxed detection regions are defined and modeled in Section~\ref{sec:fre:reg:char}. In Section~\ref{sec:opt:prec}, the optimal $M$-QAM precoder design problems for total and spatial peak power minimization are formulated and transformed into standard convex forms. We devise an interior point-based algorithm in Section~\ref{alg:ipm} to solve them. In Section~\ref{sec:sim}, we evaluate the proposed methods and compare them versus the benchmark scheme through simulations. Finally, the conclusions are made in Section~\ref{sec:con}. 
\subsection{Notation}
Upper-case and lower-case bold-faced letters are used to denote matrices and column (unless otherwise mentioned) vectors, respectively. The superscripts $(\cdot)^T$, $(\cdot)^*$, and $(\cdot)^H$
represent transpose, conjugate, and Hermitian.
${{\bf{I}}_{N \times N}}$ denotes an $N$ by $N$ identity matrix, ${\bf{0}}_{N \times 1}$ is the all zero $N$ by $1$ vector, ${\bf{E}}_k$ has one unit-valued element on the $k$-th diagonal entry with the rest of the elements being zero, ${\widetilde {\bf{E}}_k}$ has two unit-valued elements on the $k$-th and $(N_t+k)$-th diagonal entries with the rest of the elements being zero, $\rm diag(\bf{a})$ denotes a diagonal matrix where the elements of the vector $\bf{a}$ are its diagonal entries, ${\bf{a}} \circ {\bf{b}}$ is the element-wise Hadamard product,  $\|\cdot\|$ is the $l^2$ norm, and $|\cdot|$ represents the absolute value of a scalar. ${\mathop{\rm Re}\nolimits} \left(  \cdot  \right)$, ${\mathop{\rm Im}\nolimits} \left(  \cdot  \right)$, and $\arg \left(  \cdot  \right)$ represent the real valued part, imaginary valued part, and phase of a complex number, respectively. 
\section{Signal and System Model}   \label{sec:sys:mod}
Let us consider a directional modulation transmitter, denoted by $T$, having $N_t$ antennas that communicates with a receiver, denoted by $R$, equipped by $N_r$ antennas using $M$-QAM modulation. A sample of the received baseband signal, $\bf{y}$, at $R$ is 
\begin{align}
& \bf{y} = {\bf{H}}{\bf{w}} + {\bf{n}}, 
\label{eqn:eve:rec:dm}
\end{align}
where $\bf{y}$ is an $N_r \times 1$ vector denoting the received signals by $R$, ${\bf{H}} = {\left[ {{{\bf{h}}_{1}},...,{{\bf{h}}_{n}},...,{{\bf{h}}_{N_r}}} \right]^T}$ is an $N_r \times N_t$ matrix denoting the channel from $T$ to $R$, ${{\bf{h}}_{n}}$ is an $N_t \times 1$ vector containing the channel coefficients from all the transmitter antennas to the $n$-th antenna of $R$, and $\bf{w}$ is the vector containing the weights of radio frequency (RF) chains, which is the design variable in this work. The random variable ${{\bf{n}}}\sim\mathcal{CN}({\bf{0}}, \sigma _{}^2  {{\bf{I}}_{N_r^\Q \times N_r^\Q}} )$ denotes the additive white Gaussian noise at $R$, where $\mathcal{CN}$ denotes a complex and circularly symmetric random variable. The vector ${\bf{s}} = \left[ {{s_1},...,{s_n},...,{s_{N_r}}} \right]$ contains the $\mathit{M}$-QAM symbols to be communicated between $T$ and $R$ using directional modulation technology, the elements of ${{\bf{H}}}{\bf{w}} = {\left[ {{s_{{1}}^{'}},..., {s_n^{'}},..., {s_{N_r}^{'}}} \right]^T}$ are the induced $\mathit{M}$-QAM symbols on the antennas of $R$ where ${s_n^{'}}$ is the induced $\mathit{M}$-QAM symbol on the $n$-th antenna of $R$. To detect the symbols, $R$ can apply conventional detectors on each receiving antenna. 

In this work, a single carrier is used to communicate symbols over a narrow band channel. Since the transmit precoder is designed such that the received signals have the desired amplitude and phase, a simple multiple-antenna receiver is considered without processing the received signals.

In the next section, we characterize the extended and relaxed regions of the mentioned $M$-QAM constellations.
\section{Characterization of Extended and \\ Relaxed Detection Regions}   \label{sec:fre:reg:char}
\begin{figure*}[]
	\centering
    \captionsetup[subfigure]{font=scriptsize,labelfont=scriptsize}
	\begin{subfigure}[b]{0.3\textwidth}
		\raisebox{5pt}{\includegraphics[width=\textwidth]{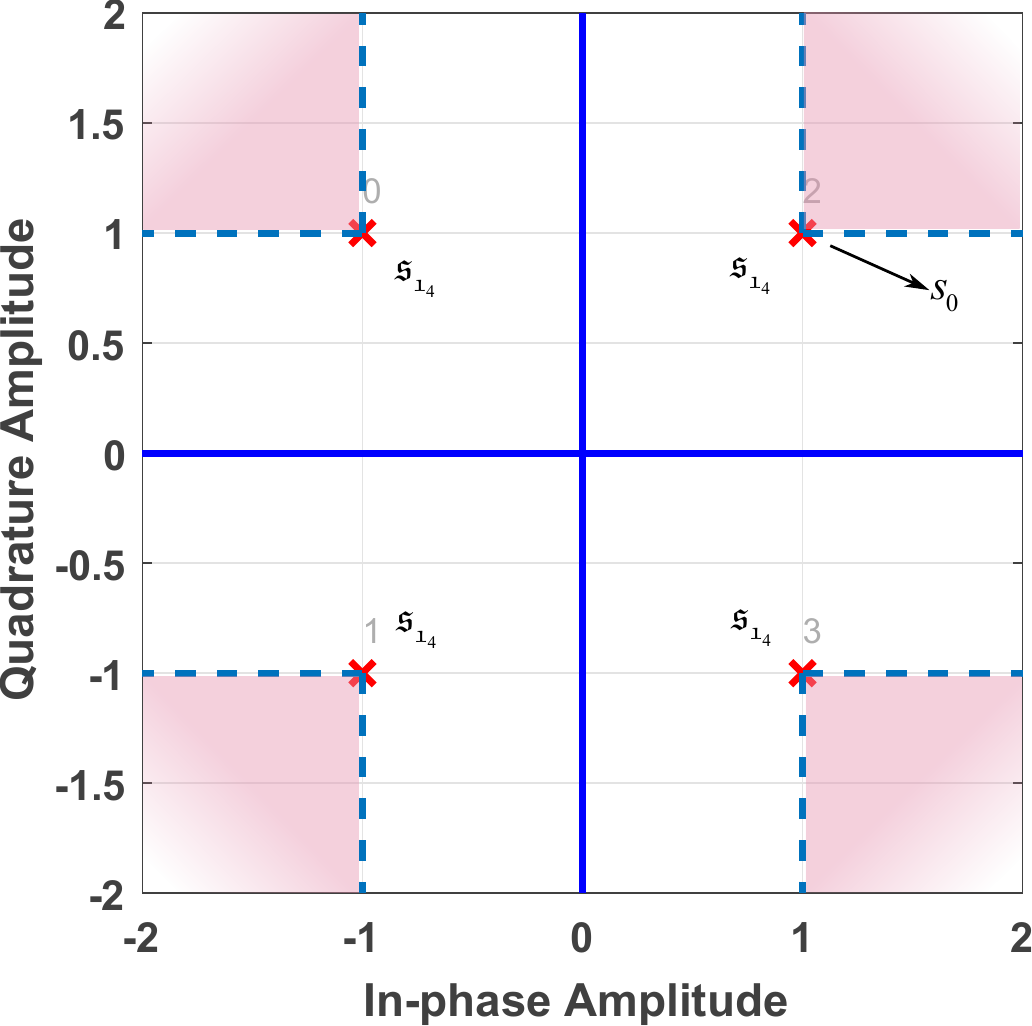}}
		\caption{Extended detection regions for $4$-QAM constellation.}
		\label{subfig:4_QAM}
	\end{subfigure}
	\begin{subfigure}[b]{0.3\textwidth}
		\includegraphics[width=\textwidth]{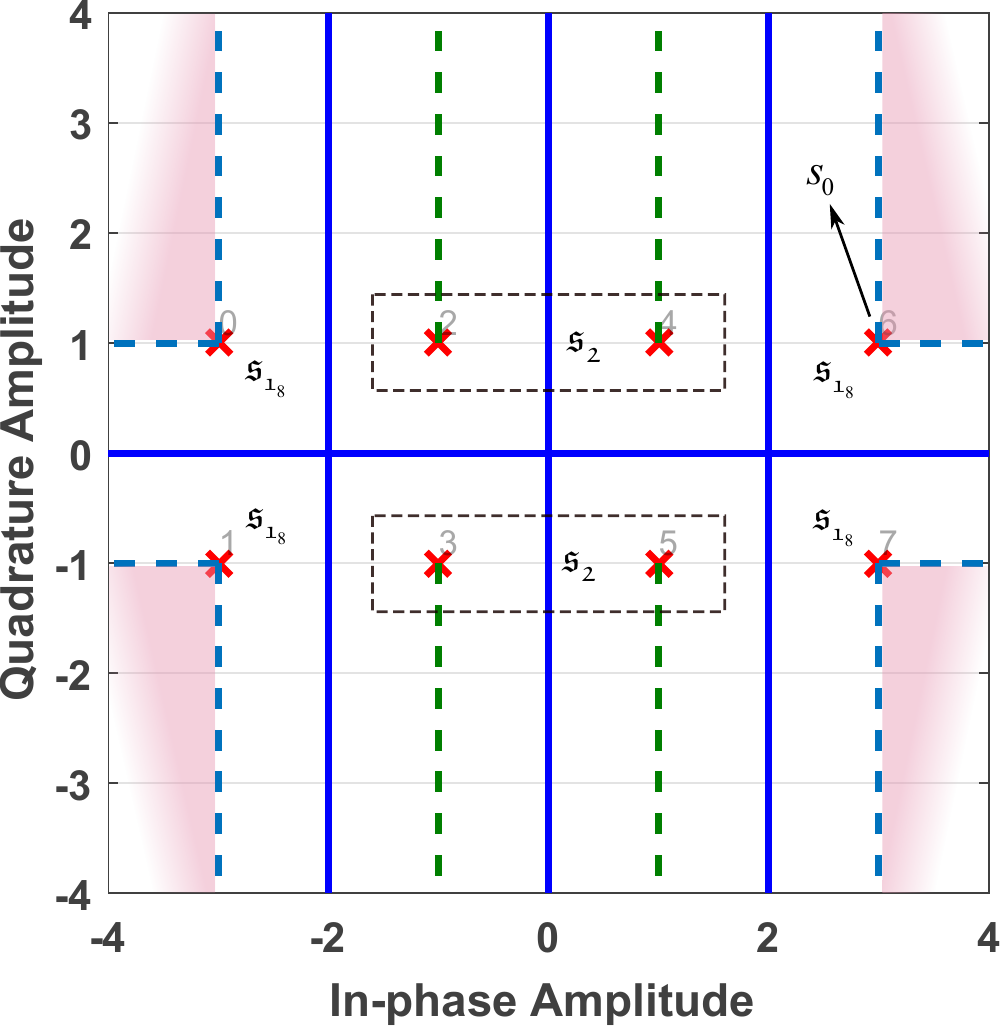}
		\caption{Extended detection regions for $8$-QAM constellation.}
		\label{subfig:8_QAM}
	\end{subfigure}
	\begin{subfigure}[b]{0.3\textwidth}
		\includegraphics[width=\textwidth]{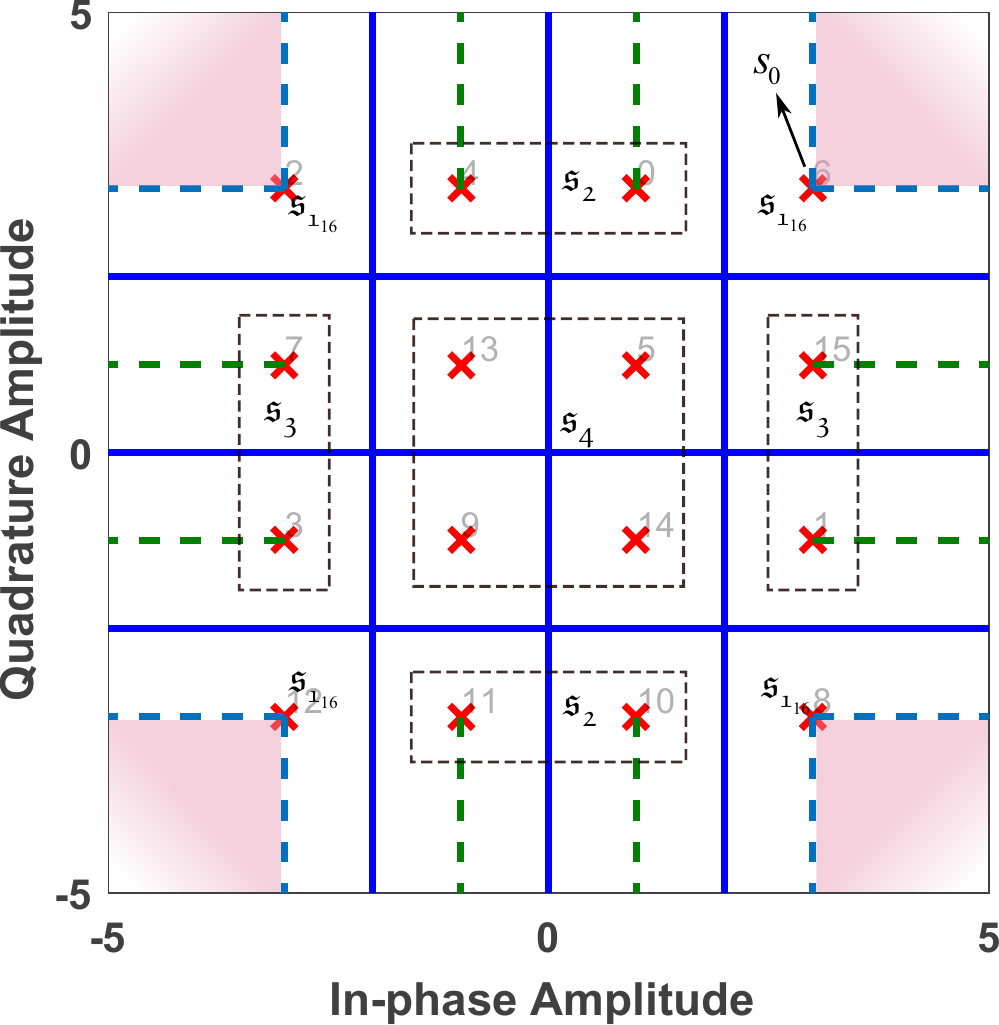}
		\caption{Extended detection regions for $16$-QAM constellation.}
		\label{subfig:16_QAM}
	\end{subfigure}
	\begin{subfigure}[b]{0.3\textwidth}
	\includegraphics[width=\textwidth]{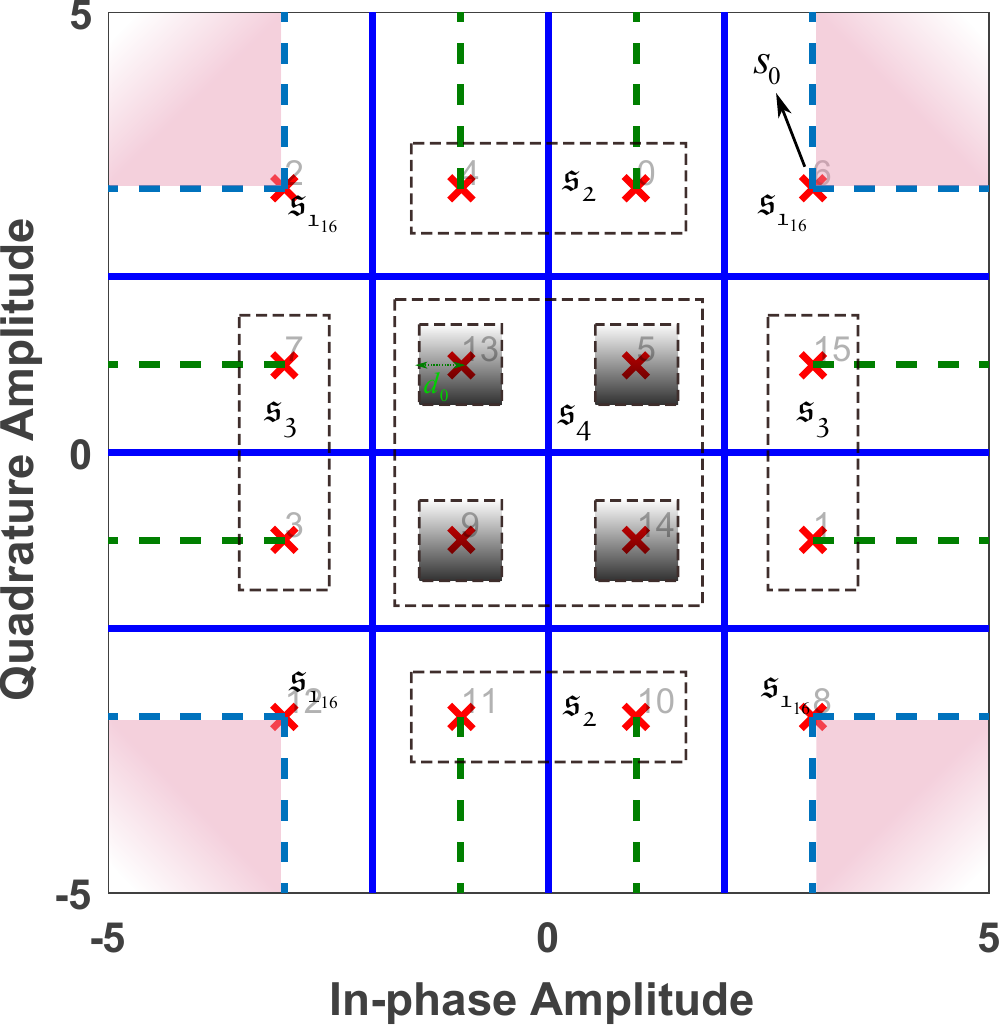}
    \caption{Extended and relaxed detection regions for $16$-QAM constellation.}
		\label{subfig:16_QAM_relax}
	\end{subfigure}
	\begin{subfigure}[b]{0.3\textwidth}
	\raisebox{15pt}{\includegraphics[width=\textwidth]{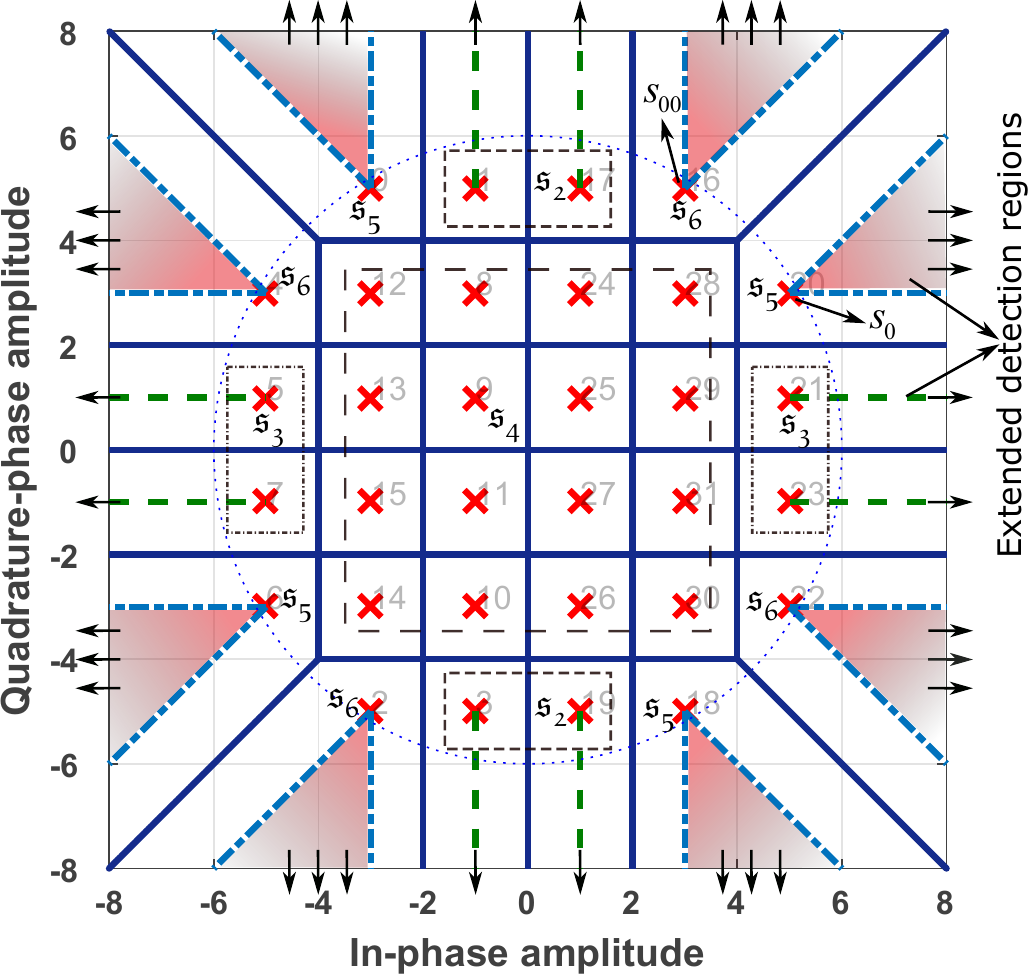}}
	\caption{Extended detection regions for $32$-QAM constellation.}
	\label{subfig:32_QAM}
	\end{subfigure}
	\begin{subfigure}[b]{0.3\textwidth}
	\raisebox{4pt}{\includegraphics[width=\textwidth]{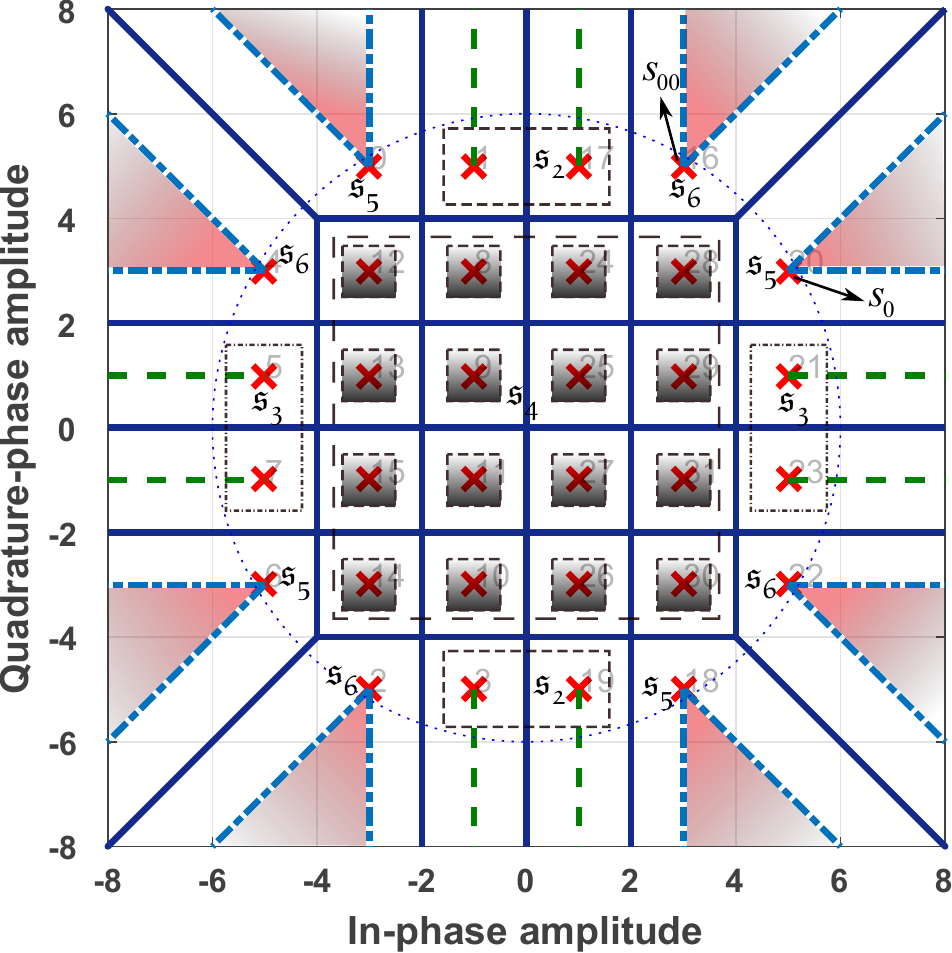}}
    \caption{Extended and relaxed detection regions for $32$-QAM constellation.}
	\label{subfig:32_QAM_relax}
	\end{subfigure}
	\caption{Characterization of extended and relaxed detection regions for $M$-QAM constellations. The numbers show the decimal equivalent of the gray code for each symbol. The extended detection regions are shown by dashed lines and solid regions for $M=4,8,16,32$.}
	\label{fig:M_QAM_cases}
\end{figure*}
In this part, we geometrically characterize the extended detection regions for $M=4,8,16,32$ constellations and relaxed detection regions for $M=16,32$ constellations. To do so, we derive analytical expressions which concisely describe these extended and relaxed detection regions. The extended detection regions are shown by solid areas and dashed lines in Fig.~\ref{fig:M_QAM_cases}. As one can see, the extended detection region is an area in which the constellation point can be placed while keeping the distance to other constellations points more than or equal to the standard distance. Here, the Euclidean distance among the constellation points of conventional $M$-QAM modulation is considered as the standard distance. We divide each constellation into multiple sets, as illustrated in Figures~\ref{subfig:4_QAM} to~\ref{subfig:32_QAM}, and continue to analytically model the extended and relaxed detection region of each set. 
\subsection{The Case of $M=4$}
Consider $s_n$ as a symbol to be communicated with the $n$-th antenna of the receiver. Modeling the received signal on the $n$-th antenna as ${\bf{h}}_n^T{\bf{w}}$, the extended detection region for $s_n \in \mathfrak{s_{1_4}}$ in the first quadrant of Fig.~\ref{subfig:4_QAM} can be modeled as 
\begin{align}
&{\rm{Re}}\left( {{\bf{h}}_n^T{\bf{w}}} \right) \ge \sqrt \gamma  {\rm{Re}}\left( {{s_n}} \right), \,\, {\rm{Im}}\left( {{\bf{h}}_n^T{\bf{w}}} \right) \ge \sqrt \gamma  {\rm{Im}}\left( {{s_n}} \right),
\label{eqn:s14:4QAM:1stq}	
\end{align}
where $\gamma$ is the minimum required amplification for the induced symbol at the receiving antennas. The value of $\gamma$ is derived from the required SNR value in dB at the receiver as $\gamma  = {10^{SNR/10}}$. The extended detection region defined in~\eqref{eqn:s14:4QAM:1stq} cannot be used in other quadrants due to change in the sign of real and imaginary parts of $s_n$. {\rc For example, if ${\rm{Re}}\left( {{s_n}} \right)$ is negative, then the absolute value of ${\rm{Re}}\left( {{\bf{h}}_n^T{\bf{w}}} \right)$ will be smaller than the absolute value of ${\rm{Re}}\left( {{s_n}} \right)$}. {\rc To handle this issue, we introduce a generalized version of the expression~\eqref{eqn:s14:4QAM:1stq} in~\eqref{eqn:s14:4QAM} where regardless of the sign of ${\mathop{\rm Re}\nolimits} \left( {{s_n}} \right)$ and ${\mathop{\rm Im}\nolimits} \left( {{s_n}} \right)$, the required signal level at the receiving antenna is satisfied.} Hence, the general expression to characterize the extended detection region for 
$s_n \in \mathfrak{s_{1_4}}$, in Fig.~\ref{subfig:4_QAM} is 
\begin{align}
&{\rm{Re}}\left( {{s_n}} \right){\rm{Re}}\left( {{\bf{h}}_n^T{\bf{w}}} \right) \ge \sqrt \gamma  {\rm{R}}{{\rm{e}}^2}\left( {{s_n}} \right), 
\nonumber\\
&{\mathop{\rm Im}\nolimits} \left( {{s_n}} \right){\rm{Im}}\left( {{\bf{h}}_n^T{\bf{w}}} \right) \ge \sqrt \gamma {{\mathop{\rm Im}\nolimits} ^2}\left( {{s_n}} \right).
\label{eqn:s14:4QAM}	
\end{align}
\subsection{The Case of $M=8$}
The extended detection region of $s_n \in \mathfrak{s_{1_8^{}}}$ for $8$-QAM constellation in Fig.~\ref{subfig:8_QAM} can be characterized in the same way as~\eqref{eqn:s14:4QAM}. Next, the upper and lower sides of $s_n \in \mathfrak{s_2}$ can be characterized, respectively, as
\begin{align}
&{\mathop{\rm Re}\nolimits} \left( {{{\bf{h}}_n^T}{\bf{w}}} \right) = \sqrt \gamma {\mathop{\rm Re}\nolimits} \left( {{s_n}} \right),\,\,
{\mathop{\rm Im}\nolimits} \left( {{{\bf{h}}_n^T}{\bf{w}}} \right) \ge \sqrt \gamma {\mathop{\rm Im}\nolimits} \left( {{s_n}} \right),
\label{eqn:u:8QAM}	
\\
&{\mathop{\rm Re}\nolimits} \left( {{{\bf{h}}_n^T}{\bf{w}}} \right) = \sqrt \gamma {\mathop{\rm Re}\nolimits} \left( {{s_n}} \right),\,\,
{\mathop{\rm Im}\nolimits} \left( {{{\bf{h}}_n^T}{\bf{w}}} \right) \le \sqrt \gamma {\mathop{\rm Im}\nolimits} \left( {{s_n}} \right).
\label{eqn:l:8QAM}	
\end{align}
The characterizations of $\mathfrak{s_2}$ in~\eqref{eqn:u:8QAM} and~\eqref{eqn:l:8QAM} can be fused to get a unified expression to describe the points in the set $\mathfrak{s_2}$ as
\begin{align}
&{\mathop{\rm Re}\nolimits} \left( {{{\bf{h}}_n^T}{\bf{w}}} \right) = \sqrt \gamma {\mathop{\rm Re}\nolimits} \left( {{s_n}} \right), \,\, {\mathop{\rm Im}\nolimits} \left( {{s_n}} \right){\mathop{\rm Im}\nolimits} \left( {{{\bf{h}}_n}{\bf{w}}} \right) \ge \sqrt \gamma {{\mathop{\rm Im}\nolimits} ^2}\left( {{s_n}} \right).
\label{eqn:ul:8-QAM}	
\end{align}
\subsection{The Case of $M=16$}
For $s_n \in \mathfrak{s_{1_{{16}^{}}}}$ of $16$-QAM constellation in Fig.~\ref{subfig:16_QAM}, the extended detection region is characterized using~\eqref{eqn:s14:4QAM}. Also, for $s_n \in \mathfrak{s_2}$ in Fig.~\ref{subfig:16_QAM}, the characterization is the same as~\eqref{eqn:ul:8-QAM}. The right-hand side and left-hand side extended detection regions of $s_n \in \mathfrak{s_3}$ can be characterized, respectively, as  
\begin{align}
{\mathop{\rm Re}\nolimits} \left( {{{\bf{h}}_n^T}{\bf{w}}} \right) \ge \sqrt \gamma {\mathop{\rm Re}\nolimits} \left( {{s_n}} \right),\,\, 
{\mathop{\rm Im}\nolimits} \left( {{{\bf{h}}_n^T}{\bf{w}}} \right) = \sqrt \gamma {\mathop{\rm Im}\nolimits} \left( {{s_n}} \right),
\label{eqn:s1:r}	
\\
{\mathop{\rm Re}\nolimits} \left( {{{\bf{h}}_n^T}{\bf{w}}} \right) \le \sqrt \gamma {\mathop{\rm Re}\nolimits} \left( {{s_n}} \right),\,\, 
{\mathop{\rm Im}\nolimits} \left( {{{\bf{h}}_n^T}{\bf{w}}} \right) = \sqrt \gamma {\mathop{\rm Im}\nolimits} \left( {{s_n}} \right),
\label{eqn:s1:l}	
\end{align}
which can be compressed into
\begin{align}
&{\mathop{\rm Re}\nolimits} \left( {{s_n}} \right){\mathop{\rm Re}\nolimits} \left( {{{\bf{h}}_n^T}{\bf{w}}} \right) \ge \sqrt \gamma {{\mathop{\rm Re}\nolimits} ^2}\left( {{s_n}} \right), {\mathop{\rm Im}\nolimits} \left( {{{\bf{h}}_n^T}{\bf{w}}} \right) = \sqrt \gamma{\mathop{\rm Im}\nolimits} \left( {{s_n}} \right).
\label{eqn:rl}	
\end{align}
In the case $s_n \in \mathfrak{s_4}$, the points can be characterized in the following two ways.
\subsubsection{Fixed detection region characterization for $s_n \in \mathfrak{s_4}$}
In this approach, the points of $s_n \in \mathfrak{s_4}$ are considered in their own place. This satisfies the minimum standard Euclidean distance between the constellation points and does not increase the SER compared to conventional $16$-QAM. However, this modeling does not improve the power efficiency. Accordingly, the constellation points $s_n \in \mathfrak{s_4}$ can be modeled as
\begin{align}
{\mathop{\rm Re}\nolimits} \left( {{\bf{h}}_n^T{\bf{w}}} \right) = \sqrt \gamma {\mathop{\rm Re}\nolimits} \left( {{s_n}} \right),\,\,
{\mathop{\rm Im}\nolimits} \left( {{\bf{h}}_n^T{\bf{w}}} \right) = \sqrt \gamma {\mathop{\rm Im}\nolimits} \left( {{s_n}} \right).
\label{eqn:s4:16QAM}	
\end{align}
\subsubsection{Relaxed detection region characterization for $s_n \in \mathfrak{s_4}$} 
\label{subsubsec:rel:16}
The points of $s_n \in \mathfrak{s_4}$ can be placed within a square-shaped detection region. This approach is illustrated in Fig.~\ref{subfig:16_QAM_relax} with gray squares. This design results in a further power consumption reduction, however, it may increase the SER since the minimum Euclidean distance between the received constellation points as in conventional $M$-QAM does not hold anymore. Relaxed detection region is modeled as 
\begin{align}
&\sqrt \gamma {\rm{Re}}\left( {{s_n}} \right) - {d_0} \le {\rm{Re}}\left( {{\bf{h}}_n^T{\bf{w}}} \right) \le \sqrt \gamma {\rm{Re}}\left( {{s_n}} \right) + {d_0},
\nonumber\\
&\sqrt \gamma {\rm{Re}}\left( {{s_n}} \right) - {d_0} \le {\rm{Im}}\left( {{\bf{h}}_n^T{\bf{w}}} \right) \le \sqrt \gamma {\rm{Im}}\left( {{s_n}} \right) + {d_0},
\label{eqn:s4:16QAM:rel}	
\end{align}
where $d_0$ is the Euclidean distance between the edge of the relaxed region and the constellation point as shown in Fig.~\ref{subfig:16_QAM_relax}.

Choosing the proper value of $d_0$ depends on the target metric such as total consumed power and SER. As an approach to derive the optimal value of $d_0$, we can perform a one dimensional search over $d_0$ to maximize the goodput~\cite{multi:adaptive:TWC} over the total consumed power defined as 
\begin{align}
\eta  = \frac{{R_s\left( {1 - SER} \right)}}{{{{\left\| {\bf{w}} \right\|}^2}}},
\label{eqn:gp}	
\end{align}
where $R_s$ is in bits per symbol.
We quantify $\eta$ in Section~\ref{sec:sim} to find its optimal value.
\subsection{The Case of $M=32$}
The extended detection regions for $s_n \in \mathfrak{s_5}$ and $s_n \in \mathfrak{s_6}$ in the first quadrant of Fig.~\ref{subfig:32_QAM} are characterized as
\begin{align}
&3 \sqrt \gamma   \le {\rm{Im}}\left( {{\bf{h}}_n^T{\bf{w}}} \right) \le {\rm{Re}}\left( {{\bf{h}}_n^T{\bf{w}}} \right) - 2\sqrt \gamma ,
\label{eqn:s31}	
\\
&{\rm{Re}}\left( {{\bf{h}}_n^T{\bf{w}}} \right) + 2\sqrt \gamma   \le {\rm{Im}}\left( {{\bf{h}}_n^T{\bf{w}}} \right),{\mkern 1mu} {\rm{Re}}\left( {{\bf{h}}_n^T{\bf{w}}} \right) \ge 3\sqrt \gamma.
\label{eqn:s32}	
\end{align}
To model the extended detection region for $s_n \in \mathfrak{s_5}$ and $s_n \in \mathfrak{s_6}$ in the other quadrants of Fig.~\ref{subfig:32_QAM}, we can rotate them so that they fall within $s_n \in \mathfrak{s_5}$ and $s_n \in \mathfrak{s_6}$ in the first quadrant. Then, we can use~\eqref{eqn:s31} and~\eqref{eqn:s32}.

The extended detection regions of $s_n \in \mathfrak{s_2}$ and $s_n \in \mathfrak{s_3}$ are modeled similarly as in~\eqref{eqn:ul:8-QAM} and~\eqref{eqn:rl}, respectively. For $s_n \in \mathfrak{s_4}$, the fixed points are modeled as ~\eqref{eqn:s4:16QAM} and the relaxed points are modeled as~\eqref{eqn:s4:16QAM:rel}. The relaxed detection regions for $s_n \in \mathfrak{s_4}$ of $32$-QAM modulation are shown in Fig.~\ref{subfig:32_QAM_relax} using gray squares.

In the next section, we design the optimal symbol-level precoders for $M$-QAM directional modulation transmitter.
\section{Directional Modulation Precoder Design}   \label{sec:opt:prec}
In this part, we design optimal $M$-QAM directional modulation precoders using the developed characterized extended and relaxed detection regions of Section~\ref{sec:fre:reg:char}. We formulate the optimal design problems in Sections~\ref{subsec:opt:pre:min:pow} and~\ref{subsec:opt:pre:LPAPR} to minimize the transmit power and spatial peak power, respectively.  
\subsection{Precoder Design: Transmit Power Minimization}   \label{subsec:opt:pre:min:pow}
In this part, we formulate and design the optimal $M$-QAM MIMO directional modulation precoder when the objective is to minimize the total transmission power while satisfying the required SNR at the receiving antennas.

The design problem for $4$-PSK case can be written as
\begin{align}
& \mathop {\min }\limits_{{\bf{w}}} \,\, {\left\| {\bf{w}} \right\|^2}
\nonumber\\
& \,\, \text{s.t.}   \,\,\, {\rm{Re}}\left( {s_{n_{1}^{}}} \right){\rm{Re}}\left( {{\bf{h}}_{n_1^{}}^T{\bf{w}}} \right) \ge \sqrt \gamma   {\rm{R}}{{\rm{e}}^2}\left( {s_{n_{1}^{}}} \right), {s_{n_{1}^{}}} \in \mathfrak{s_{1_4}}
\nonumber\\
&  \qquad  {\mathop{\rm Im}\nolimits} \left( {s_{n_{1}^{}}} \right){\rm{Im}}\left( {{\bf{h}}_{n_1^{}}^T{\bf{w}}} \right) \ge \sqrt \gamma {{\mathop{\rm Im}\nolimits} ^2}\left( {s_{n_{1}^{}}} \right).
\label{eqn:dm:4-QAM:pow:min}
\end{align}
We can cast the optimal design for $8$-QAM modulation as
\begin{subequations}
	\begin{align}
	& \mathop {\min }\limits_{{\bf{w}}} \,\,\, {\left\| {\bf{w}} \right\|^2}
	\nonumber\\
	&  \,\, \text{s.t.}   \,\,\, {\mathop{\rm Re}\nolimits} \left( {s_{n_{{1}}^{}}} \right){\rm{Re}}\left( {{\bf{h}}_{{n_{{1}}^{}}}^T{\bf{w}}} \right) \!\!  \ge \!\! \sqrt \gamma  {{\mathop{\rm Re}\nolimits} ^2}\left( {{s_{{n_{{1}}^{}}}}} \right), 
	\\
	& \qquad {\mathop{\rm Im}\nolimits} \left( {{s_{{n_{{1}}^{}}}}} \right){\rm{Im}}\left( {{\bf{h}}_{{n_{{1}}^{}}}^T{\bf{w}}} \right) \!\! \ge \!\! \sqrt \gamma {{\mathop{\rm Im}\nolimits} ^2}\left( {{s_{{n_{{1}}^{}}}}} \right) \!,s_{n_{1}^{}} \in \mathfrak{s_{1_8}}
	\label{subeq:dm:8-QAM1}
	\\
	&             \qquad  {\rm{Re}}\left( {{\bf{h}}_{{n_2^{}}}^T{\bf{w}}} \right) = \sqrt \gamma {\rm{Re}}\left( {{s_{{n_2^{}}}}} \right), 
	\\
	& \qquad {\rm{Im}}\left( {{s_{{n_2^{}}}}} \right){\rm{Im}}\left( {{{\bf{h}}_{{n_2^{}}}}{\bf{w}}} \right) \ge \sqrt \gamma {\rm{I}}{{\rm{m}}^2}\left( {{s_{{n_2^{}}}}} \right), s_{n_2^{}} \in \mathfrak{s_{2}}
	\label{subeq:dm:8-QAM3}			
	\end{align}
	\label{eqn:dm:8-QAM:pow:min}%
\end{subequations}
Next, we can formulate the optimal precoder design problem for $16$-QAM as follows
\begin{subequations}
	\begin{align}
	& \mathop {\min }\limits_{{\bf{w}}} \,\,\, {\left\| {\bf{w}} \right\|^2}
	\nonumber\\
	&  \,\, \text{s.t.}   \,\,\, {\rm{Re}}\left( {s_{n_{{1}}^{}}} \right){\rm{Re}}\left( {{\bf{h}}_{n_1^{}}^T{\bf{w}}} \right) \!\! \ge \!\! \sqrt \gamma {\rm{R}}{{\rm{e}}^2}\left( {s_{n_{{1}}^{}}} \right), {s_{n_{{1}}^{}}} \in \mathfrak{s_{1_{16}}}
	\\
	&             \qquad  {\mathop{\rm Im}\nolimits} \left( {s_{n_{{1}}^{}}} \right){\rm{Im}}\left( {{\bf{h}}_{n_1^{}}^T{\bf{w}}} \right) \ge \sqrt \gamma {{\mathop{\rm Im}\nolimits} ^2}\left( {s_{n_{{1}}^{}}} \right),
	\\
	&             \qquad    {\rm{Re}}\left( {{\bf{h}}_{{n_2^{}}}^T{\bf{w}}} \right) = \sqrt \gamma {\rm{Re}}\left( {{s_{{n_2^{}}}}} \right), s_{n_{2}^{}} \in \mathfrak{s_{2}}
	\label{subeq:dm:16-QAM21}
	\\
	&             \qquad    {\rm{Im}}\left( {{s_{{n_2^{}}}}} \right){\rm{Im}}\left( {{{\bf{h}}_{{n_2^{}}}}{\bf{w}}} \right) \ge \sqrt \gamma {\rm{I}}{{\rm{m}}^2}\left( {{s_{{n_2^{}}}}} \right), 
	\label{subeq:dm:16-QAM31}	
	\\
	&             \qquad    {\rm{Re}}\left( {{s_{{n_3^{}}}}} \right){\rm{Re}}\left( {{\bf{h}}_{{n_3^{}}}^T{\bf{w}}} \right) \ge \sqrt \gamma {\rm{R}}{{\rm{e}}^2}\left( {{s_{{n_3^{}}}}} \right), s_{n_3^{}} \in \mathfrak{s_{3}}
	\label{subeq:dm:16-QAM41}
	\\
	&             \qquad    {\rm{Im}}\left( {{\bf{h}}_{{n_3^{}}}^T{\bf{w}}} \right) = \sqrt \gamma {\rm{Im}}\left( {{s_{{n_3^{}}}}} \right).
	\label{subeq:dm:16-QAM511}
	\\
	&             \qquad 	{\rm{Re}}\left( {{\bf{h}}_{{n_4^{}}}^T{\bf{w}}} \right) = \sqrt \gamma {\rm{Re}}\left( {{s_{{n_4^{}}}}} \right), s_{n_4^{}} \in \mathfrak{s_{4}}
	\label{subeq:dm:16-QAM521}
	\\
	&             \qquad 	{\rm{Im}}\left( {{\bf{h}}_{{n_4^{}}}^T{\bf{w}}} \right)= \sqrt \gamma {\rm{Im}}\left( {{s_{{n_4^{}}}}} \right).
	\label{subeq:dm:16-QAM61}
	\end{align}
	\label{eqn:dm:16-QAM:pow:min}%
\end{subequations}
In the case of relaxed detection region design for inner points, the constraints~\eqref{subeq:dm:16-QAM521} and~\eqref{subeq:dm:16-QAM61} are replaced by~\eqref{eqn:s4:16QAM:rel}. Finally, the precoder design problem for $32$-QAM is defined as
\begin{subequations}
	\begin{align}
	& \mathop {\min }\limits_{{\bf{w}}} \,\,\, {\left\| {\bf{w}} \right\|^2}
	\nonumber\\
	&  \,\, \text{s.t.}   \,\,\,\,\,\, {\mathop{\rm Re}\nolimits} \left( {{{\bf{h}}_{n_2^{}}^T}{\bf{w}}} \right) =\sqrt \gamma {\mathop{\rm Re}\nolimits} \left( {{s_{n_2^{}}}} \right),\, s_{n_2^{}} \in \mathfrak{s_2}
	\label{subeqn:dm:32-QAM111}
	\\
	&             \qquad \,\,\, 	{\rm{Im}}\left( {{s_{{n_2^{}}}}} \right){\rm{Im}}\left( {{{\bf{h}}_{{n_2^{}}}}{\bf{w}}} \right) \ge \sqrt \gamma {\rm{I}}{{\rm{m}}^2}\left( {{s_{{n_2^{}}}}} \right),
	\label{subeqn:dm:32-QAM121}
	\\
	&             \qquad \,\,\,  	{\mathop{\rm Re}\nolimits} \left( {{s_{n_3^{}}}} \right){\mathop{\rm Re}\nolimits} \left( {{{\bf{h}}_{n_3^{}}^T}{\bf{w}}} \right) \! \ge \! \sqrt \gamma {{\mathop{\rm Re}\nolimits}^2}\left( {{s_{n_3^{}}}} \right),\, s_{n_3^{}} \in \mathfrak{s_3}
	\label{subeqn:dm:32-QAM131}
	\\
	&             \qquad \,\,\,		{\mathop{\rm Im}\nolimits} \left( {{{\bf{h}}_{n_3^{}}^T}{\bf{w}}} \right) =\sqrt \gamma {\mathop{\rm Im}\nolimits} \left( {{s_{n_3^{}}}} \right),
	\label{subeqn:dm:32-QAM141}
	\\
	&             \qquad \,\,\,  {\mathop{\rm Re}\nolimits} \left( {{\bf{h}}_{n_4^{}}^T{\bf{w}}} \right) =\sqrt \gamma {\mathop{\rm Re}\nolimits} \left( {{s_{n_4^{}}}} \right),\, s_{n_4^{}} \in \mathfrak{s_4}
	\label{subeqn:dm:32-QAM151}
	\\
	&             \qquad \,\,\,  	{\mathop{\rm Im}\nolimits} \left( {{\bf{h}}_{n_4^{}}^T{\bf{w}}} \right) =\sqrt \gamma {\mathop{\rm Im}\nolimits} \left( {{s_{n_4^{}}}} \right),
	\label{subeqn:dm:32-QAM161}
	\\
	&             \qquad \,\,\,  3 \sqrt \gamma \!\! \le \!\! {\rm{Im}}\left( {\widetilde {\bf{h}}_{{n_5^{}}}^T{\bf{w}}} \right) \!\! \le \!\! {\rm{Re}}\left( {\widetilde {\bf{h}}_{{n_5^{}}}^T{\bf{w}}} \right) \!\!-\!\! 2 \sqrt \gamma, s_{n_5^{}} \in \mathfrak{s_5} 
	\label{subeqn:dm:32-QAM171}
	\\
	&             \qquad \,\,\,  {\rm{Re}}\left( {\widetilde {\bf{h}}_{{n_6^{}}}^T{\bf{w}}} \right) + 2 \sqrt \gamma  \le {\rm{Im}}\left( {\widetilde {\bf{h}}_{{n_6^{}}}^T{\bf{w}}} \right), \, s_{n_6^{}} \in \mathfrak{s_6}
	\label{subeqn:dm:32-QAM181}
	\\
	&             \qquad \,\,\,  {\rm{Re}}\left( {\widetilde {\bf{h}}_{{n_6^{}}}^T{\bf{w}}} \right) \ge 3 \sqrt \gamma, 
	\label{subeqn:dm:32-QAM191}				
	\end{align}
	\label{eqn:dm:32-QAM1:pow:min}%
\end{subequations}
where $\widetilde {\bf{h}}_{{n_5^{}}}^T = {\bf{h}}_{{n_5^{}}}^T{e^{i{\varphi _{{n_5^{}}}}}}$, $\widetilde {\bf{h}}_{{n_6^{}}}^T = {\bf{h}}_{{n_6^{}}}^T{e^{i{\varphi _{{n_6^{}}}}}}$, ${{\varphi _{{n_5^{}}}}}$ is the phase difference between $s_0$, shown in Fig.~\ref{subfig:32_QAM}, and $s_{n_5^{}} \in \mathfrak{s_5}$, ${{\varphi _{{n_6^{}}}}}$ is the phase difference between $s_{00}$, shown in Fig.~\ref{subfig:32_QAM}, and $s_{n_6^{}} \in \mathfrak{s_6}$. In the case of relaxed design for inner constellation points, we can replace the constraints~\eqref{subeqn:dm:32-QAM151} and~\eqref{subeqn:dm:32-QAM161} by~\eqref{eqn:s4:16QAM:rel}. 

We transform~\eqref{eqn:dm:32-QAM1:pow:min} into a standard convex form when considering fixed and relaxed detection regions for the points $s_{n_4^{}} \in \mathfrak{s_4}$. A similar approach can be applied to the design problems in~\eqref{eqn:dm:4-QAM:pow:min} to~\eqref{eqn:dm:16-QAM:pow:min} in order to get standard convex forms.
 
After applying a series of algebraic operations on~\eqref{eqn:dm:32-QAM1:pow:min}, shown in Appendix~\ref{app:trans}, the transformed $32$-QAM design problem becomes
\begin{align}
& \mathop {\min }\limits_{{\bf{w}}} \,\,\, {\left\| {\bf{w}} \right\|^2}  \,\, \text{s.t.}   \,\,\,\,\,\, {\bf{Aw}} \ge {\bf{a}}, \, {\bf{Bw}} = {\bf{b}},
\label{eqn:pre:32:qam:fix:final}
\end{align}
where $\bf{A}$, $\bf{a}$, $\bf{B}$, and $\bf{b}$ are defined in~\eqref{eqn:def}. In the case of relaxed detection region design for the points $s_{n_4^{}} \in \mathfrak{s_4}$, we can reach a problem with the same format as in~\eqref{eqn:pre:32:qam:fix:final} where $\bf{A}$, $\bf{a}$, $\bf{B}$ and $\bf{b}$ are defied as in~\eqref{eqn:def:rel}. As we see,~\eqref{eqn:pre:32:qam:fix:final} is a convex linearly constrained quadratic programming problem. We propose an interior point algorithm in Section~\ref{alg:ipm} to solve it.
\subsection{Precoder Design: Spatial Peak Power Minimization}   \label{subsec:opt:pre:LPAPR}
In this part, we design the optimal $M$-QAM directional modulation precoders for $M=4,8,16,32$ aiming at keeping the power output of each power amplifier as low as possible. We pursuit this design while satisfying the required SNR at the receiving antennas. Note that in another approach we can minimize the transmission power while keeping the output power of each RF chain below a specific value and satisfying the required SNR at the receiving antennas.

The optimal spatial peak power minimization precoder design problem for $4$-QAM modulation is written as
\begin{align}
& \mathop {\min }\limits_{{\bf{w}}} \,\, \mathop {\max }\limits_{k = 1,...,{N_t}} \,\,{{\bf{w}}^H}{{\bf{E}}_k}{\bf{w}} \,\,\,\,\,\,\,\, \text{s.t.}   \,\,\,   {\Omega _{M-QAM}},
\label{eqn:dm:M-QAM}
\end{align}
where ${\Omega _{M-QAM}}$ is the constellation-specific constraint set which can be found for $M=4,8,16,32$ in~\eqref{eqn:dm:4-QAM:pow:min},~\eqref{eqn:dm:8-QAM:pow:min},~\eqref{eqn:dm:16-QAM:pow:min}, and~\eqref{eqn:dm:32-QAM1:pow:min}, respectively. In the case of relaxed detection region design for $s_{n_4^{}} \in \mathfrak{s_4}$, the related constraints, defined in~\eqref{eqn:dm:16-QAM:pow:min} and~\eqref{eqn:dm:32-QAM1:pow:min}, are replaced by the constraints in~\eqref{eqn:s4:16QAM:rel}. 

Here, we proceed with transforming~\eqref{eqn:dm:M-QAM} for $M=32$ into a standard form. A similar approach can be applied to the design problems for $M=4,8,16$. First, we move the objective of~\eqref{eqn:dm:M-QAM} to the constraint by introducing the auxiliary variable $z$ as
\begin{align}
& \mathop {\min }\limits_{{\bf{w}},z} \,\, z  \,\,\,\,\,\,\,\, \text{s.t.}   \,\,\,  {{\bf{w}}^H}{{\bf{E}}_k}{\bf{w}} \le z, \,\,\, {\Omega _{M-QAM}} \,\,\, \forall \, k = 1,...,{N_t}
\label{eqn:dm:M-QAM2}
\end{align}
To transform~\eqref{eqn:dm:M-QAM2} for $M=32$, we can apply a similar process used in Appendix~\ref{app:trans} for~\eqref{eqn:dm:32-QAM1:pow:min} to get
\begin{align}
	& \mathop {\min }\limits_{{\bf{w}},z} \,\,\, z \,\,\,\,\,\,\,\, \text{s.t.}   \,\,\, {{\bf{w}}^T}{{\widetilde {\bf{E}}_k}}{\bf{w}} \le z, \,\,\, {\bf{Aw}} \ge {\bf{a}}, 
	\nonumber\\
	& \qquad \qquad \qquad {\bf{Bw}} = {\bf{b}}, \,\,\, \forall \, k = 1,...,{N_t} 
\label{eqn:pre:gen4}
\end{align}
where $\bf{A}$, $\bf{a}$, $\bf{B}$ and $\bf{b}$ are as in~\eqref{eqn:def}. As we see,~\eqref{eqn:pre:gen4} is a convex problem which can be solved efficiently. If we consider the relaxed detection region design for $s_{n_4^{}} \in \mathfrak{s_4}$, we get a similar problem as in~\eqref{eqn:pre:gen4} where $\bf{A}$, $\bf{a}$, $\bf{B}$, and $\bf{b}$ are the same as~\eqref{eqn:def:rel}. In the next section, we propose an interior point path-following algorithm to solve~\eqref{eqn:pre:gen4}.
\section{Interior Point path-following algorithm} \label{alg:ipm}
Here, we devise an interior point path-following algorithm~\cite{Nesterov1994} to solve~\eqref{eqn:pre:gen4}. First, we create self-concordant barriers and include an approximation of the inequality constraints of~\eqref{eqn:pre:gen4} in the objective function and drive the barrier generated family as ${{{F}}_t} = tf + {F}$ where $f$ is the objective in~\eqref{eqn:pre:gen4} and $F$ is the summation of self-concordant barrier functions. As $t \to  + \infty$, $F$ becomes a better approximation of the constraints. Here, $t$ is multiplied by $\mu$ in each iteration step. 

Now, we define the self-concordant barrier function for each inequality constraint of~\eqref{eqn:pre:gen4}. Using the summation rule for barriers~\cite{Nesterov1994}, the $N_t$-self-concordant barrier function for the first group of constraints in~\eqref{eqn:pre:gen4} becomes 
\begin{align}
{F_1}\left( {{\bf{w}},z} \right) =  - \sum\nolimits_{k = 1}^{N_t} {\ln \left( {z - {{\bf{w}}^T}{{\widetilde{\bf{E}}}_k}{\bf{w}}} \right)}. 
\label{eqn:sc:C1}	
\end{align}
Using the summation rule again, the self-concordant barrier for the second constraint of~\eqref{eqn:pre:gen4} is given by 
\begin{align}
{F_2}\left( {\bf{w}} \right) =  - \sum\nolimits_{k = 1}^{{r_{\bf{A}}^{}}} {\ln \left( { - {a_k} + {\bf{a}}_k{\bf{w}}} \right)}, 
\label{eqn:sc:C2}	
\end{align}
where ${r_{\bf{A}}^{}}$ is the number of rows in $\bf{A}$, ${{\bf{a}}_k}$ is the $k_{th}$ row of $\bf{A}$ and ${{a_k}}$ is the $k_{th}$ element of $\bf{a}$. The function in~\eqref{eqn:sc:C2} is a ${r_{\bf{A}}^{}}$-self-concordant barrier. 

Incorporating~\eqref{eqn:sc:C1} and~\eqref{eqn:sc:C2} in the objective, we get an equality constraint minimization as
\begin{align}
& \mathop {\min }\limits_{{\bf{w}},z,t} \,\,\, tz + F\left( {{\bf{w}},z} \right) \,\,\,  \text{s.t.}    \qquad {\bf{Bw}} = {\bf{b}},
\label{eqn:eqcon}
\end{align}
where $F\left( {{\bf{w}},z} \right) = {F_1}\left( {{\bf{w}},z} \right) + {F_2}\left( {\bf{w}} \right)$ is a $N_t + {r_{\bf{A}}^{}}$-self-concordant barrier function. For a specific value of $t$, we can write the Karush-Kuhn-Tucker (KKT) conditions for~\eqref{eqn:eqcon} as
	\begin{align}
    & {\nabla _{{\bf{w}},z}}\left[ {tz + F\left( {{{\bf{w}}^{\star}},z^{\star}} \right)} \right] + {{\bf{B}}^T}\boldsymbol{\lambda}^{\star} = 0, \,\, {\bf{B}}{{\bf{w}}^{\star}} = {\bf{b}},
\label{eqn:kkt}
	\end{align}
%
where ${{\bf{w}}^{\star}}$, $z^{\star}$, and $\boldsymbol{\lambda}^{\star}$ are the optimal primal and dual variables. We can use the Newton method to solve the system of non-linear equations in~\eqref{eqn:kkt} to find the descent directions. For starting points ${\bf{w}}$, $z$, and $\boldsymbol\lambda$, we linearize the KKT equations in~\eqref{eqn:kkt} in terms of the descent directions $\Delta {\bf{w}}$ and $\Delta z$ as
\begin{subequations}
	\begin{align}
	&{\nabla _{{\bf{w}},z}}\left[ {t\left( {z + \Delta z} \right) + F\left( {{\bf{w}} + \Delta {\bf{w}},z + \Delta z} \right)} \right] + {{\bf{B}}^T}\boldsymbol{\lambda} 
	\nonumber\\
	&={\nabla _{{\bf{w}},z}}f + {\nabla ^2}_{{\bf{w}},z}f\left[ {\begin{array}{*{20}{l}}
		{\Delta {\bf{w}}}\\
		{\Delta z}
		\end{array}} \right]+ {{\bf{B}}^T}\boldsymbol{\lambda}, 
	\label{subeqn:grad:lin}
	\\
	& {\bf{B}}\left( {{\bf{w}} + \Delta {\bf{w}}} \right) = {\bf{b}} \,\, \to \,\, {\bf{B}}\Delta {\bf{w}} = {\bf{0}},
	\label{subeqn:pri:feas:lin}
	\end{align}
	\label{eqn:kkt:lin}%
\end{subequations}
where $f = tz + F\left( {{\bf{w}},z} \right)$. Using the linearized equations in~\eqref{eqn:kkt:lin}, the descent directions are found by solving
\begin{align}
\left[ {\begin{array}{*{20}{c}}
	{{\nabla ^2}_{{\bf{w}},z}f}&{{{\bf{B}}^T}}\\
	{\bf{B}}&{{{\bf{0}}_{{ (r_{\bf{B}}^{} + 1 )} \times \left( {{r_{\bf{B}}^{}} + 1} \right)}}}
	\end{array}} \right]  \left[ \begin{array}{l}
\Delta {\bf{w}}\\
\Delta z\\
\boldsymbol\lambda 
\end{array} \right]  =  - \left[ \begin{array}{l}
{\nabla _{{\bf{w}},z}}f\\
{0_{{{r_{\bf{B}}^{}}} \times 1}}
\end{array} \right]
\label{eqn:des:dir}
\end{align}
where ${r_{\bf{B}}^{}}$ shows the number of rows of $\bf{B}$. After getting $\Delta {\bf{w}}$ and $\Delta z$ from~\eqref{eqn:des:dir}, we carry out the following updates
\begin{align}
{{\bf{w}}_{i + 1}} \!\! = \!\! {{\bf{w}}_i} \! + \! \alpha \left( {{{\Delta {\bf{w}}} \mathord{\left/
			{\vphantom {{\Delta {\bf{w}}} {\left\| {\Delta {\bf{w}}} \right\|}}} \right.
			\kern-\nulldelimiterspace} {\left\| {\Delta {\bf{w}}} \right\|}}} \right), \! {z_{i + 1}} \!\! = \!\! {z_i} \! +\! \alpha \left( {{{\Delta z} \mathord{\left/
			{\vphantom {{\Delta z} {\left\| {\Delta z} \right\|}}} \right.
			\kern-\nulldelimiterspace} {\left\| {\Delta z} \right\|}}} \right),
\label{eqn:update:w:t}
\end{align}
where $\alpha$ is found using the backtracking line search~\cite{Dennis1996NMf}. Backtracking line search method guarantees that, first, we avoid small decrease in $f$ relative to the step length and second, the steps are not too small relative to the rate of decrease in $f$. Next, we drive the values of ${\nabla _{{\bf{w}},z}}f$ and ${{\nabla ^2}_{{\bf{w}},z}f}$ as
\begin{subequations}
	\begin{align}
{\nabla _{{\bf{w}},z}}f =& t \left[ \begin{array}{l}
{{\bf{0}}_{2{N_t}}}\\
1
\end{array} \right] - \sum\limits_{k = 1}^{{N_t}} {\frac{1}{{\left( {{{\bf{w}}^T}{{\widetilde {\bf{E}}}_k}{\bf{w}} - z} \right)}}\left[ \begin{array}{l}
	2{{\bf{E}}_i}{\bf{w}}\\
	- 1
	\end{array} \right]} 
\nonumber\\
&- \sum\limits_{k = 1}^{{r_{\bf{A}}^{}}} {\frac{1}{{ - {{\bf{a}}_k} {\bf{w}} + {a_k}}}\left[ \begin{array}{l}
	- {\bf{a}}_k^T\\
	0
	\end{array} \right]}, 
	\\
	 \nabla _{{\bf{w}},z}^2f =& \sum\limits_{k = 1}^{{N_t}} {\frac{1}{{{{\left( {{{\bf{w}}^T}{{\widetilde {\bf{E}}}_k}{\bf{w}} - z} \right)}^2}}}\left[ \begin{array}{l}
		2{{\bf{E}}_k}{\bf{w}}\\
		- 1
		\end{array} \right]{{\left[ \begin{array}{l}
				2{{\bf{E}}_k}{\bf{w}}\\
				- 1
				\end{array} \right]}^T}}
				\nonumber\\
			&+ \sum\limits_{k = 1}^{{N_t}} {\frac{1}{{ - \left( {{{\bf{w}}^T}{{\widetilde {\bf{E}}}_k}{\bf{w}} - z} \right)}}} \left[ {\begin{array}{*{20}{c}}
		{2{{\bf{E}}_k}}&{{{\bf{0}}_{2{N_t} \times 1}}}\\
		{{{\bf{0}}_{1 \times 2{N_t}}}}&0
		\end{array}} \right]
	\nonumber\\
	&+ \sum\limits_{k = 1}^{{r_{\bf{A}}^{}}} {\frac{1}{{{{\left( { - {{\bf{a}}_k}{\bf{w}} + {a_k}} \right)}^2}}}\left[ \begin{array}{l}
		- {\bf{a}}_k^T\\
		0
		\end{array} \right]{{\left[ \begin{array}{l}
				- {\bf{a}}_k^T\\
				0
				\end{array} \right]}^T}}. 
	\label{subeqn:hessf}
	\end{align}
	\label{eqn:grad:hess}%
\end{subequations}
\algnewcommand{\algorithmicgoto}{\textbf{Go to}}%
\algnewcommand{\Goto}[1]{\algorithmicgoto~\ref{#1}}%
\begin{algorithm}[t]
	\caption{Interior point algorithm to solve~\eqref{eqn:pre:gen4}}
	\begin{algorithmic}[1]
		\State Define the $N_t + {r_{\bf{A}}^{}}$-self-concordant barrier, denoted by $F$ using~\eqref{eqn:sc:C1} and~\eqref{eqn:sc:C2};
		\State Get ${\bf{w}}_0$ and $z_0$ inside the feasible region by solving~\eqref{eqn:fea:pro}; 
		\State Obtain the values of $t_0$ and $\boldsymbol{\lambda}_0$ using~\eqref{eqn:ini:t:lambda};
		\State Set values of the constants $\mu$, ${\varepsilon_1}$ and ${\varepsilon_2}$;		
		\State Set $i_1=0$ 
		\If {${{\left( {  N_t + {r_{\bf{A}}^{}} } \right)} \mathord{\left/
					{\vphantom {{\left( {{r_1} + k} \right)} t}} \right.
					\kern-\nulldelimiterspace} t_{i_1^{}}} > {\varepsilon _1}$}
		\State Set $i_2=0$;		
		\If {$\kappa  > 2{\varepsilon_2}$}
		\State Find the descent direction $\Delta {\bf{w}}$ and $\Delta z$ using~\eqref{eqn:des:dir}\label{descent};
		\State Perform backtracking line search to find $\alpha$;
		\State Update variables as ${{\bf{w}}_{i_2^{} + 1}} = {{\bf{w}}_{i_2^{}}} + \alpha \left( {{{\Delta {\bf{w}}} \mathord{\left/
					{\vphantom {{\Delta {\bf{w}}} {\left\| {\Delta {\bf{w}}} \right\|}}} \right.
					\kern-\nulldelimiterspace} {\left\| {\Delta {\bf{w}}} \right\|}}} \right), 
					{z_{{i_2^{}} + 1}} = {z_{i_2^{}}} + \alpha \left( {{{\Delta z} \mathord{\left/
					{\vphantom {{\Delta z} {\left\| {\Delta z} \right\|}}} \right.
					\kern-\nulldelimiterspace} {\left\| {\Delta z} \right\|}}} \right)$;
		\State Set $i_2=i_2+1$;
		\Else
		\State Set ${t_{{i_1^{}} + 1}} = \mu {t_{{i_1^{}}}}$;
		\State Set ${i_1} = {i_1} + 1$;
		\EndIf
		\EndIf
	\end{algorithmic}
	\label{alg:int}
\end{algorithm}
Replacing~\eqref{eqn:grad:hess} into~\eqref{eqn:des:dir}, we drive the trajectory directions of the path-following interior point denoted by ${\bf{w}}\left( t \right)$ and $z\left( t \right)$. To have a descent direction, ${{\nabla ^2}_{{\bf{w}},z}f}$ needs to be positive definite. Considering that the trajectories ${\bf{w}}\left( t \right)$ and $z\left( t \right)$ are feasible for each value of $t$ and using relations~\eqref{eqn:grad:hess}, we can see that ${{\nabla ^2}_{{\bf{w}},z}f}$ is non-negative definite. However, it is possible that ${{\nabla ^2}_{{\bf{w}},z}f}$ becomes close to singularity, in this case, we can consider ${\nabla ^2}_{{\bf{w}},z}f + \varepsilon_0 {\bf{I}}$ instead, where $\varepsilon_0$ is sufficiently big to make ${\nabla ^2}_{{\bf{w}},z}f$ non-singular. To start the path-following interior point approach, we need an initial feasible point which satisfies the constraints~\eqref{eqn:pre:gen4}. To do so, we can solve
\begin{align}
& \mathop {\min }\limits_{{\bf{w}},z,s} \,\,\, s \,\,\, \text{s.t.}   \,\,\,\,\,\, {{\bf{w}}^T}{\widetilde {\bf{E}}_k}{\bf{w}} - z \le s, \,\,\, - {\bf{Aw}} + {\bf{a}} \le s, 
\nonumber\\
&   \qquad \qquad \qquad {\bf{Bw}} = {\bf{b}}, \,\,\, \forall \, k = 1,...,{N_t}.
\label{eqn:fea:pro}
\end{align}
To find an initial point for~\eqref{eqn:fea:pro}, we find ${\bf{w}}$ to satisfy the equality constraint, then $z$ and $s$ can be found to satisfy the first and second constraints. After this , problem~\eqref{eqn:fea:pro} can be solved using Algorithm~\ref{alg:int} to find $\bf{w}_0$ and $z_0$. The process of solving~\eqref{eqn:fea:pro} is stopped when a negative value for $s$ is found. Then, the derived value of $\bf{w}$ and $z$ can be used as initial points to solve~\eqref{eqn:pre:gen4} using the path-following interior point method. Now, it remains to find $t_0$ and ${\boldsymbol{\lambda}}_0$. We aim to find these parameters so that the optimality condition in~\eqref{eqn:kkt} is minimized. 
To pursue this, we can solve the following least squares problem
\begin{align}
\left[ {{t_0},{\boldsymbol{\lambda_0}}} \right] = \arg \mathop {\min }\limits_{t,\boldsymbol{\lambda} } {\left\| {{\nabla_{{\bf{w}},z}}\left[ {tz_0^{} + F\left( {{{\bf{w}}_0^{}},{z_0^{}}} \right)} \right] + {{\bf{B}}^T}\boldsymbol{\lambda} } \right\|_2}.
\label{eqn:ini:t:lambda}
\end{align}
As the stopping criteria, we use the Newton decrement accuracy measurement factor defined as 
\begin{align}
\kappa  = {\left[ {\begin{array}{*{20}{c}}
		{\Delta {\bf{w}}}\\
		{\Delta z}
		\end{array}} \right]^T}\nabla _{{\bf{w}},z}^2f\left[ {\begin{array}{*{20}{c}}
	{\Delta {\bf{w}}}\\
	{\Delta z}
	\end{array}} \right],
\label{eqn:New:dec}
\end{align}
Detailed steps of the devised algorithm are mentioned in Algorithm~\ref{alg:int}. The same approach can be applied to solve~\eqref{eqn:pre:32:qam:fix:final}. 

To benchmark our algorithm, we compare it with CVX optimization toolbox for parameters $N_t=N_r=5$, $100$ symbols, $100$ channel realizations, $\mu=5$, $\epsilon_1=\epsilon_2 = 6 \times 10^{-2}$. Considering ${{{\bf{w}}_{ip}}}$ and ${{{\bf{w}}_{cvx}}}$ as the solutions of our algorithm and CVX, respectively, the error percentage is calculated as ${{\left\| {{{\bf{w}}_{ip}} - {{\bf{w}}_{cvx}}} \right\|} \mathord{\left/
		{\vphantom {{\left\| {{{\bf{w}}_{ip}} - {{\bf{w}}_{cvx}}} \right\|} {\left\| {{{\bf{w}}_{cvx}}} \right\|}}} \right.
		\kern-\nulldelimiterspace} {\left\| {{{\bf{w}}_{cvx}}} \right\|}}$. Our algorithm results in $3.62\%$ and $2.80\%$ average error for optimization variables $z$ and $\bf{w}$, respectively, and is in average $0.53$ seconds faster. Since the precoder is designed for each group of symbols in directional modulation, the $0.53$ matters here.
\begin{figure*}[]
	\centering
	\begin{subfigure}[b]{0.49\textwidth}
		\includegraphics[width=\textwidth]{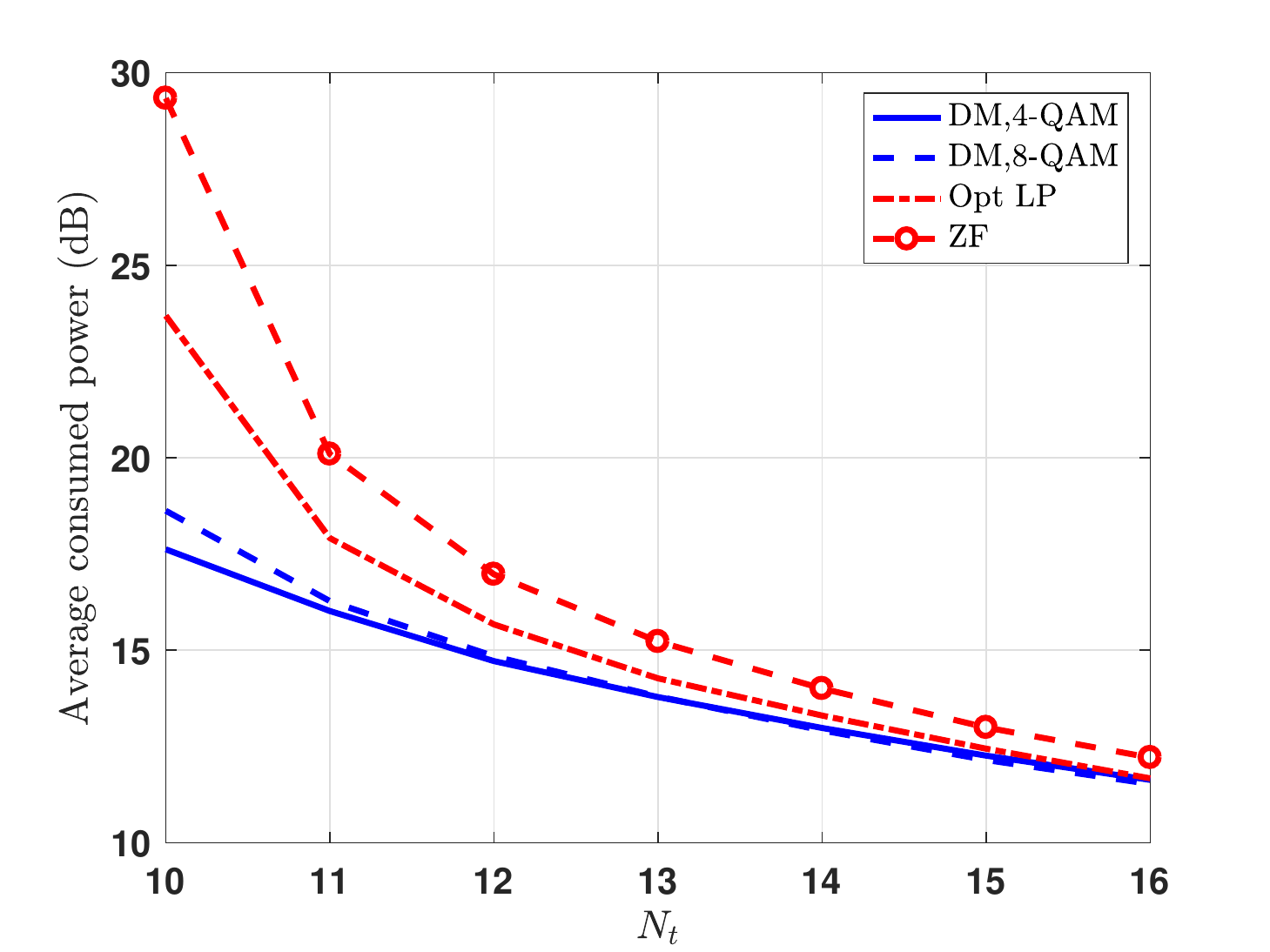}
		\caption{$\mathit{SNR}=10$ dB.}
		\label{subfig:fig:pow:ant1}
	\end{subfigure}
	\begin{subfigure}[b]{0.49\textwidth}
		\includegraphics[width=\textwidth]{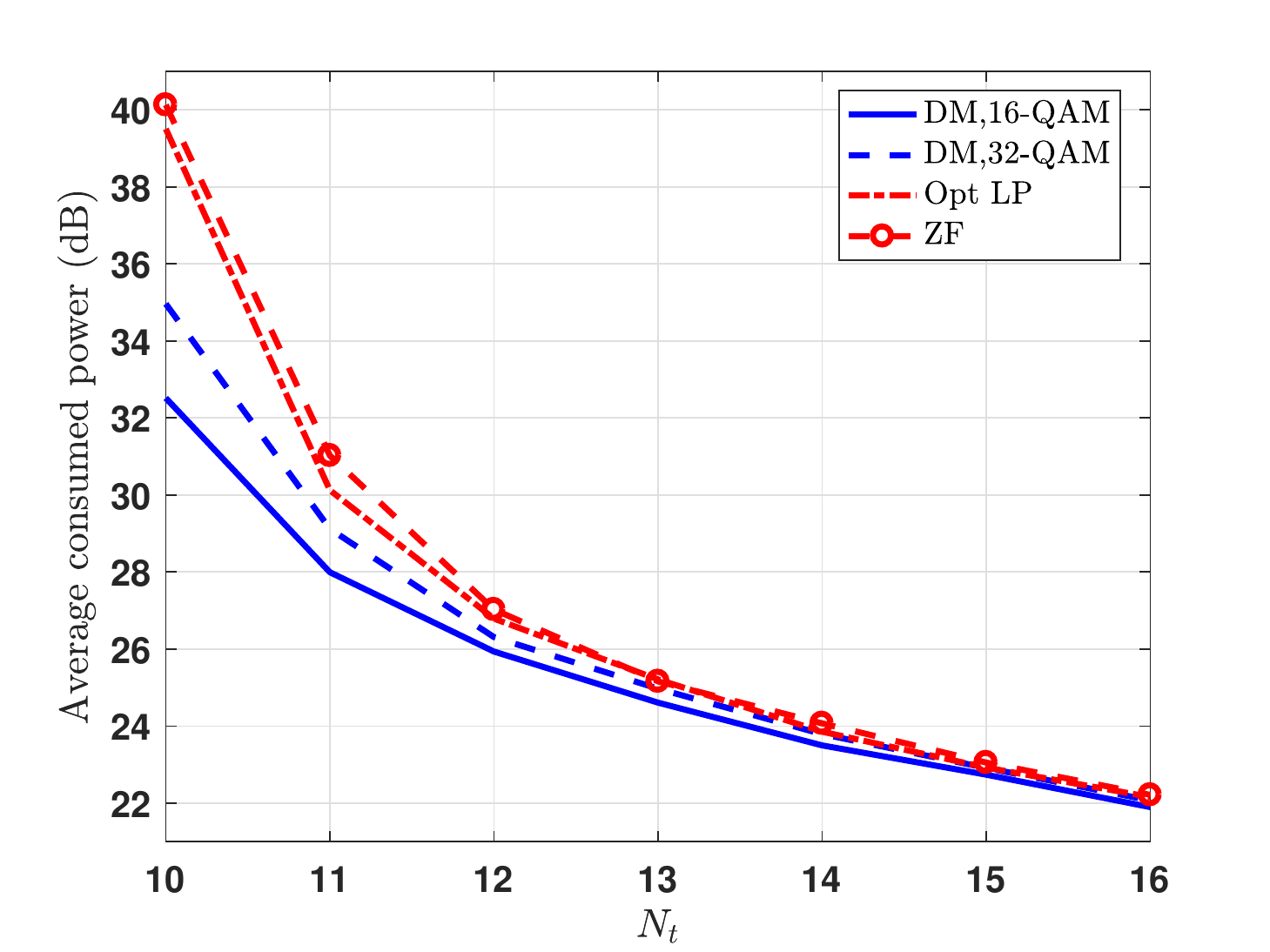}
		\caption{$\mathit{SNR}=20$ dB.}
		\label{subfig:fig:pow:ant2}
	\end{subfigure}
	\caption{Average total consumed power with respect to $N_t$ for the proposed M-QAM directional modulation precoding and the benchmark schemes when total and spatial peak power minimization designs are considered with $N_r=10$.}
	\label{fig:pow:ant}
\end{figure*}
\begin{figure*}[]
	\centering
	\begin{subfigure}[b]{0.49\textwidth}
		\includegraphics[width=\textwidth]{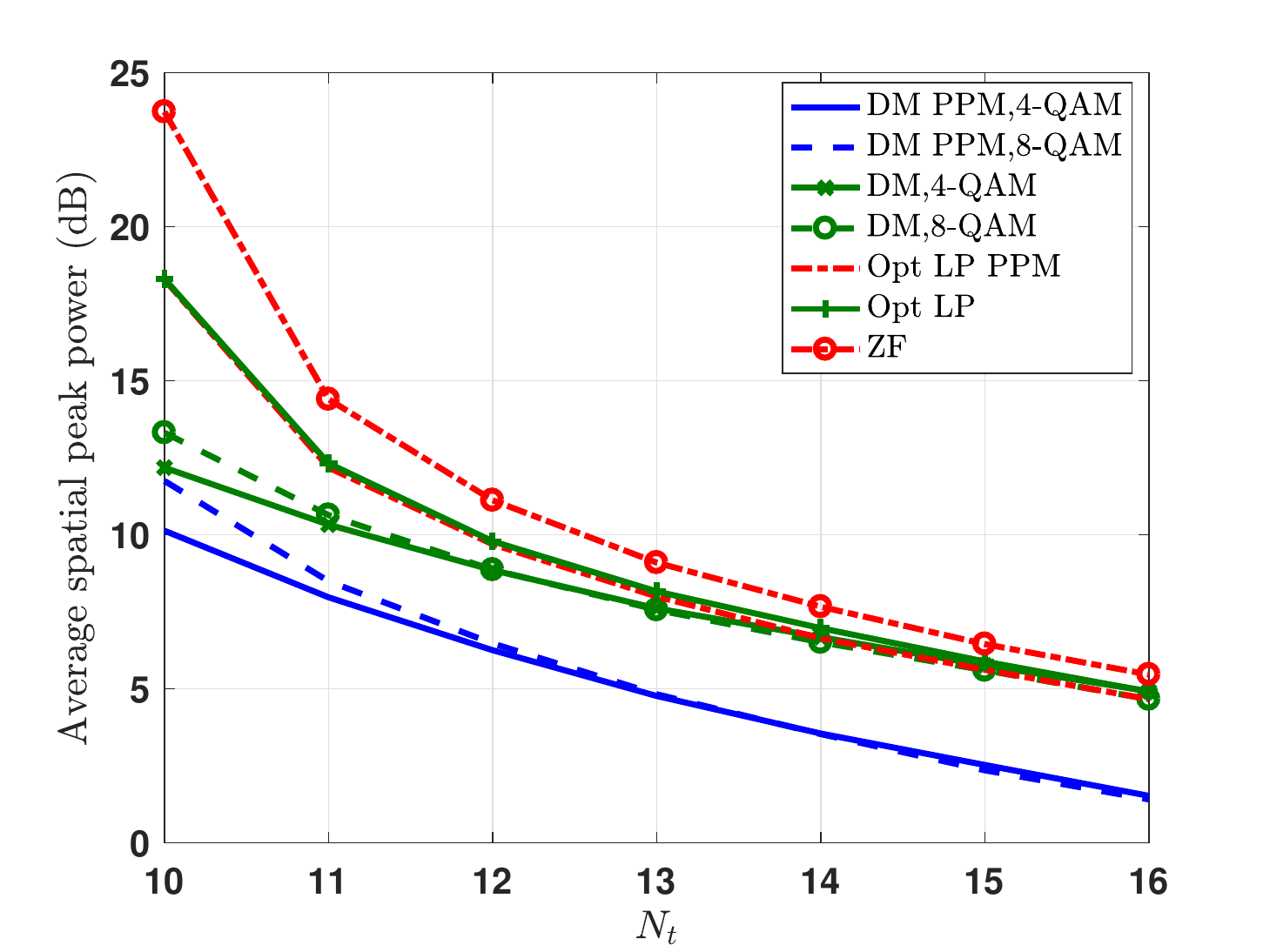}
		\caption{$\mathit{SNR}=10$ dB.}
		\label{subfig:pow_ant_max_fixed1}
	\end{subfigure}
	\begin{subfigure}[b]{0.49\textwidth}
		\includegraphics[width=\textwidth]{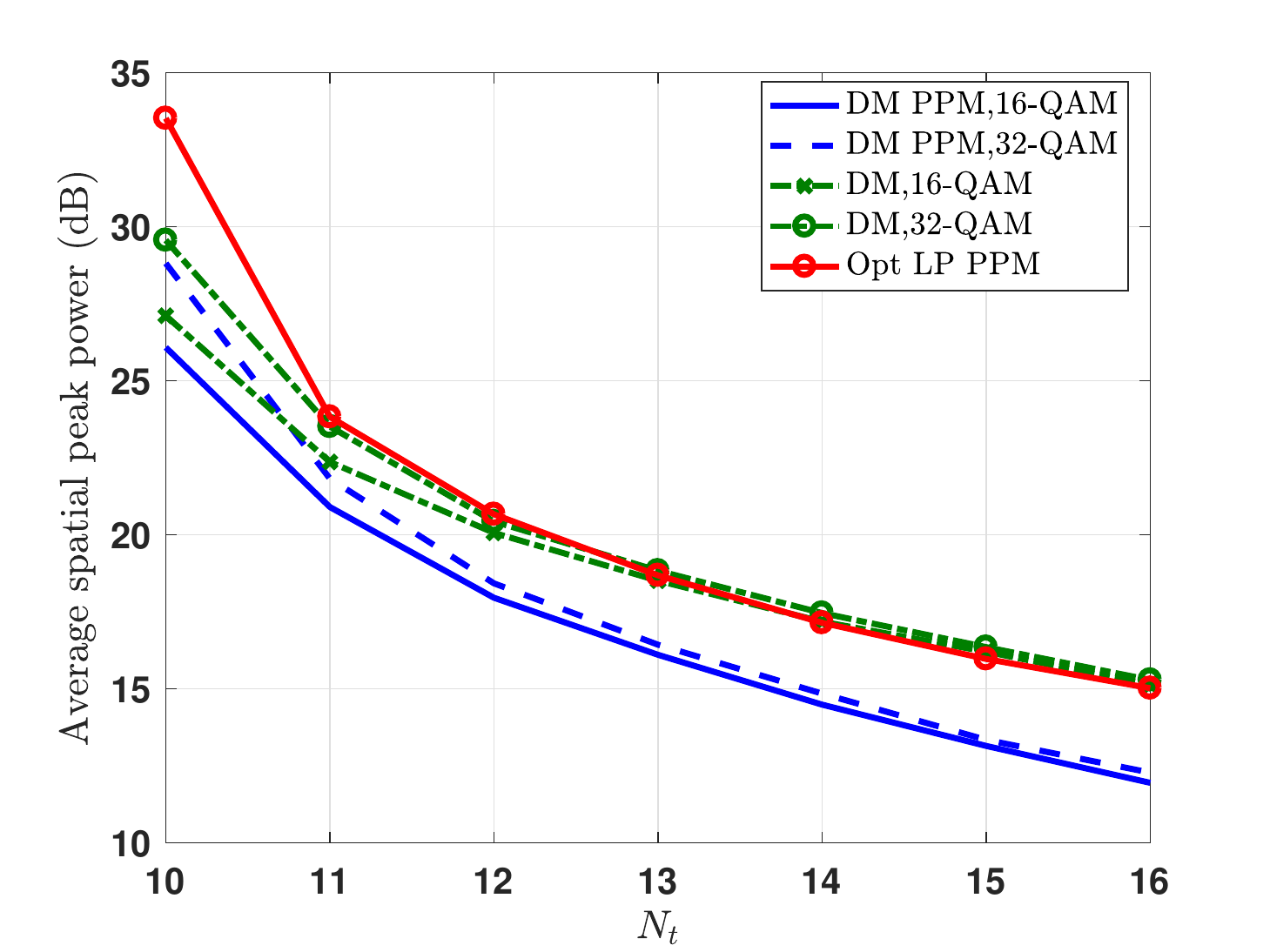}
		\caption{$\mathit{SNR}=20$ dB.}
		\label{subfig:pow_ant_max_fixed2}
	\end{subfigure}
	\caption{Average maximum peak power among the RF transmit chains with respect to $N_t$ for the proposed M-QAM directional modulation precoding and the benchmark schemes when total and spatial peak power minimization designs are considered with $N_r=10$.}
	\label{fig:pow:ant:max}
\end{figure*}
\section{Simulation Results}   \label{sec:sim}
\begin{figure*}[]
	\centering
	\begin{subfigure}[b]{0.49\textwidth}
		\includegraphics[width=\textwidth]{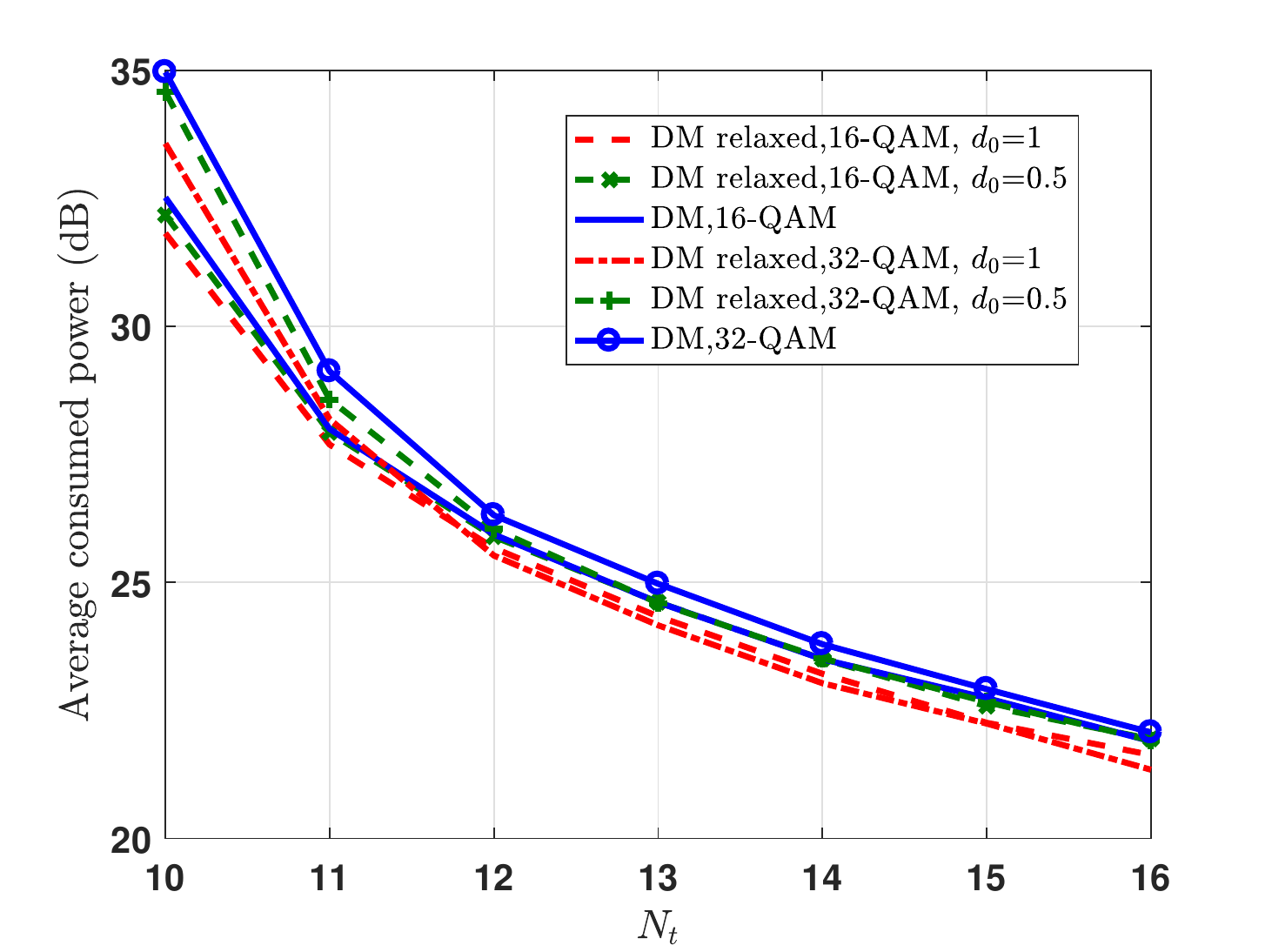}
		\caption{ }
		\label{fig:pow:ant:relaxed}
	\end{subfigure}
	\begin{subfigure}[b]{0.49\textwidth}
		\includegraphics[width=\textwidth]{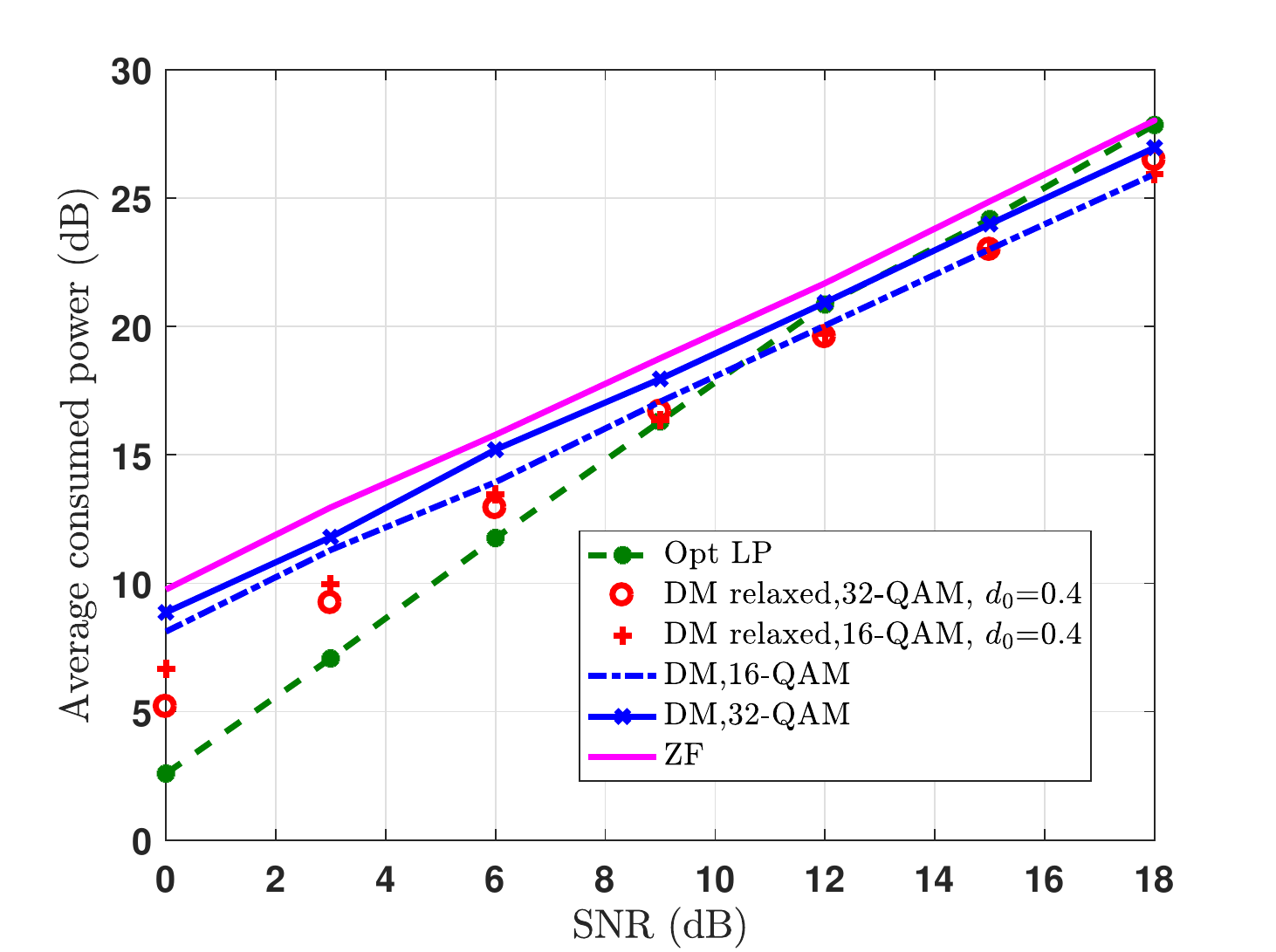}
		\caption{ }
		\label{fig:pow:snr:fixed:Nt11}
	\end{subfigure}
	\caption{Average consumed power with respect to $N_t$ and SNR for the proposed $M$-QAM directional modulation precoding scheme when fixed and relaxed detection region designs for $s_{n_4^{}} \in \mathfrak{s_4}$ are considered. Fig.~\ref{fig:pow:ant:relaxed} parameters are $N_r=10$ and $\mathit{SNR}=20$ dB. Fig.~\ref{fig:pow:snr:fixed:Nt11} parameters are $N_t=11$, $N_r=10$.}
	\label{fig:pow:snr2}
\end{figure*}
\begin{figure*}[]
	\centering
	\begin{subfigure}[b]{0.49\textwidth}
		\includegraphics[width=\textwidth]{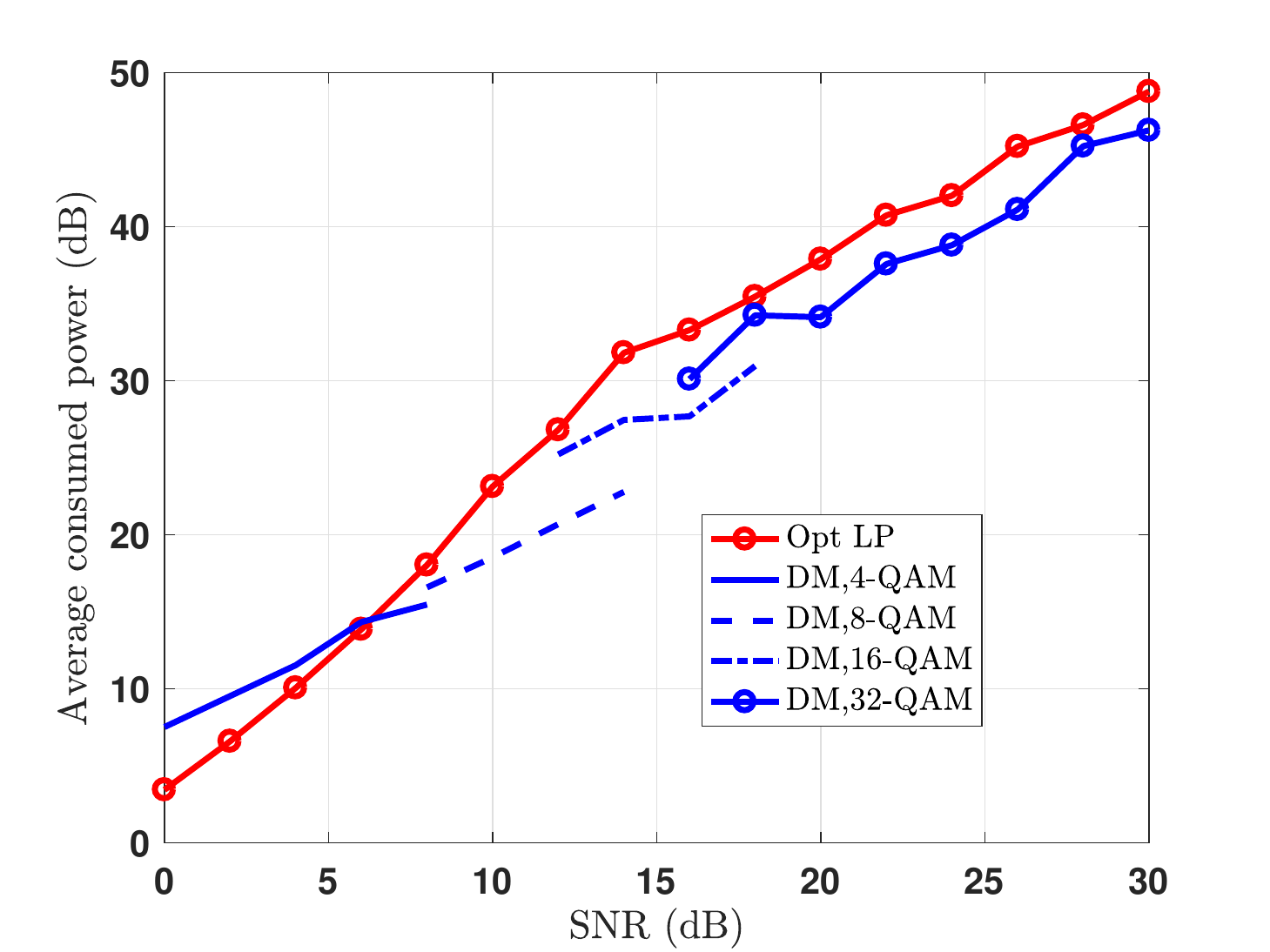}
		\caption{ }
		\label{subfig:pow_snr}
	\end{subfigure}
	\begin{subfigure}[b]{0.49\textwidth}
		\includegraphics[width=\textwidth]{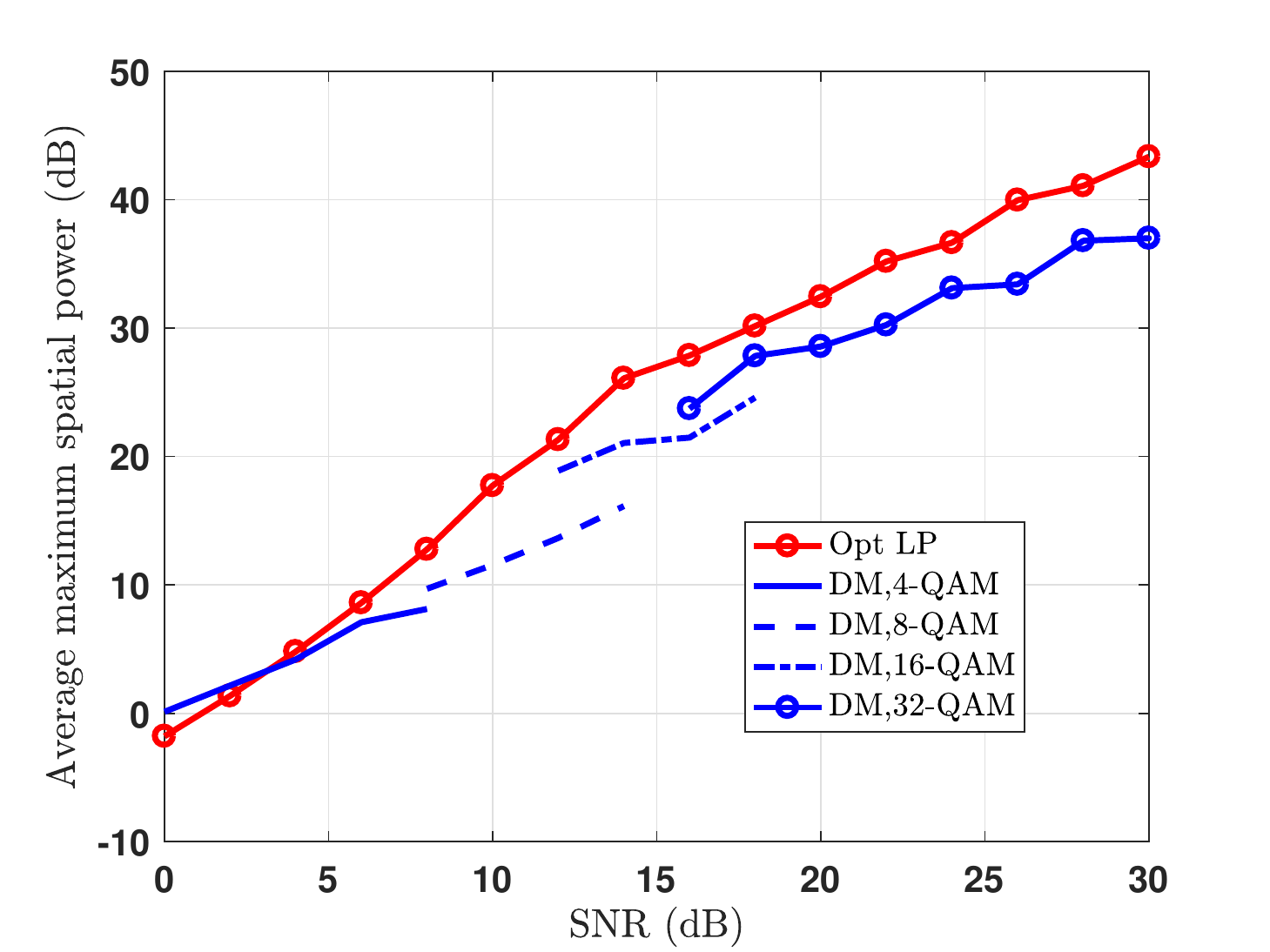}
		\caption{ }
		\label{subfig:pow_snr_max}
	\end{subfigure}
	\caption{Average total consumed and spatial peak powers with respect to the required SNR for the proposed $M$-QAM directional modulation precoding and the benchmark schemes when $N_t=N_r=10$.}
	\label{fig:pow:snr}
\end{figure*}
\begin{figure*}[]
	\centering
	\begin{subfigure}[b]{0.49\textwidth}
		\includegraphics[width=\textwidth]{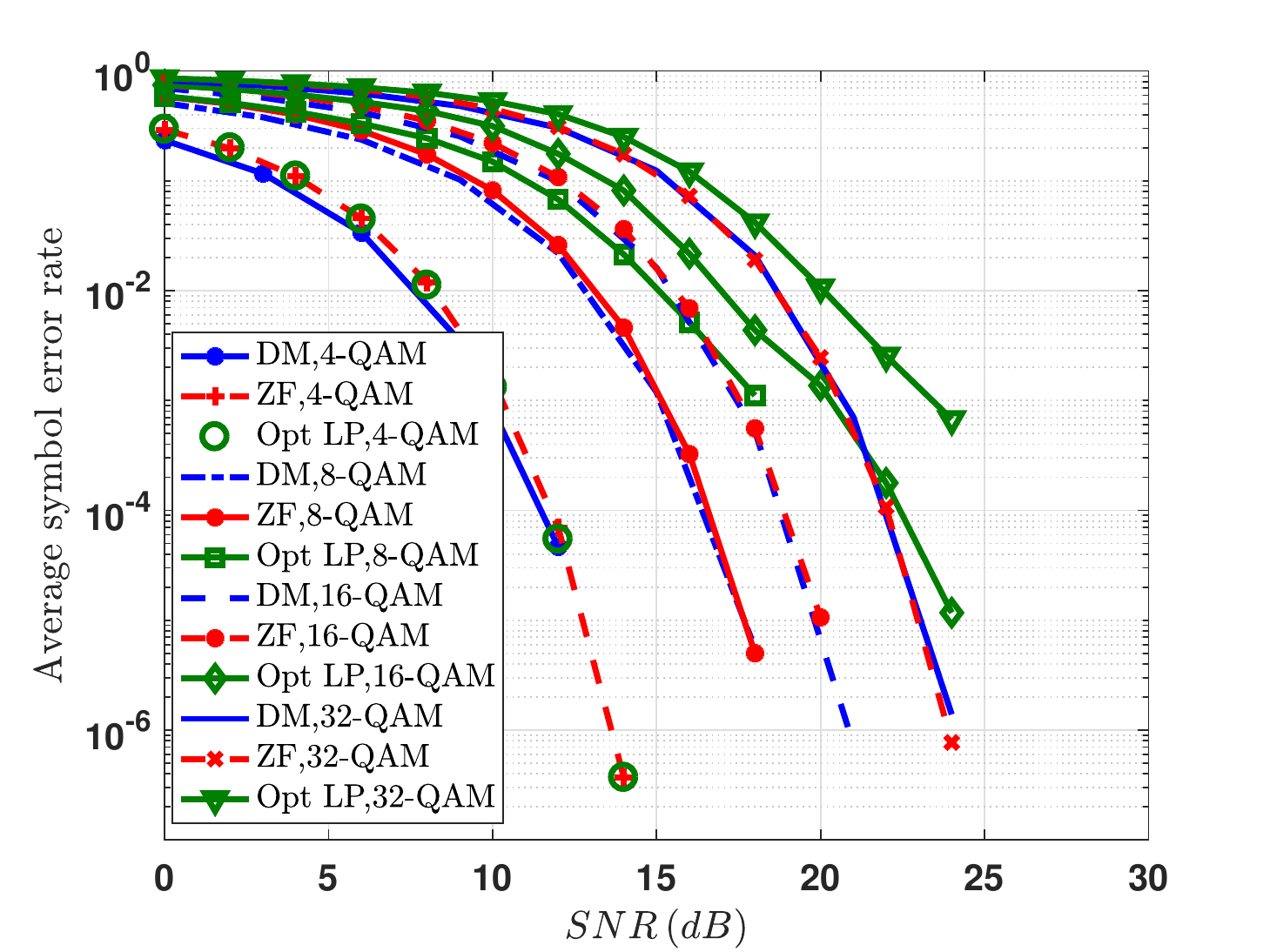}
		\caption{ }
		\label{fig:ser}
	\end{subfigure}
	\begin{subfigure}[b]{0.49\textwidth}
		\includegraphics[width=\textwidth]{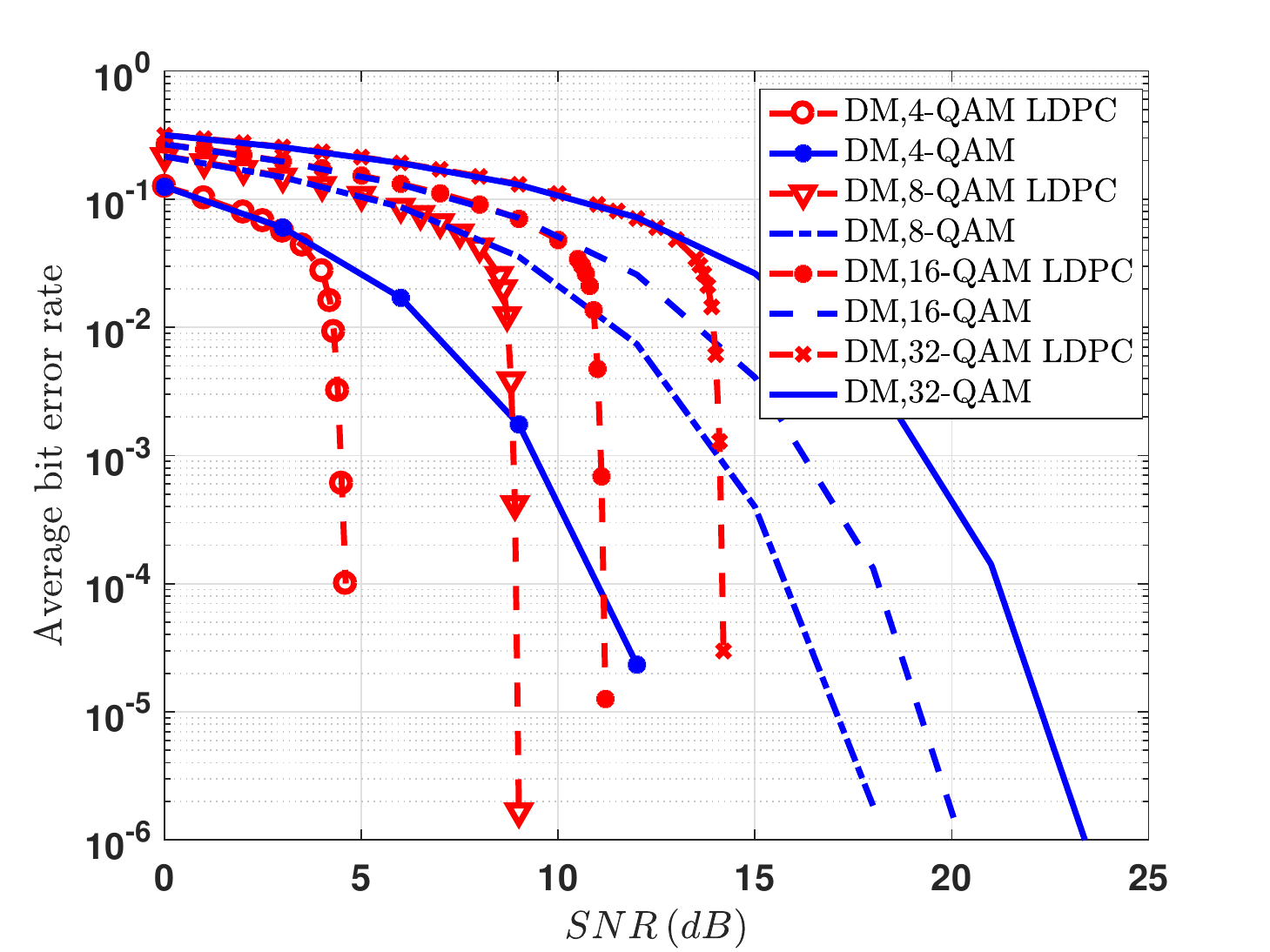}
		\caption{ }
		\label{fig:ber}
	\end{subfigure}
	\caption{Average SER and BER with respect to the required SNR at the receiver for the proposed $M$-QAM directional modulation precoding with $N_t=N_r=10$. Fig.~\ref{fig:ser} shows the comparison with benchmark schemes and Fig.~\ref{fig:ber} shows the BER of our scheme with and without LDPC forward error correction code when the code rate is $5/6$.}
	\label{fig:ser:ber}
\end{figure*}
\begin{figure}[]
	\centering
	\includegraphics[width=9.5cm]{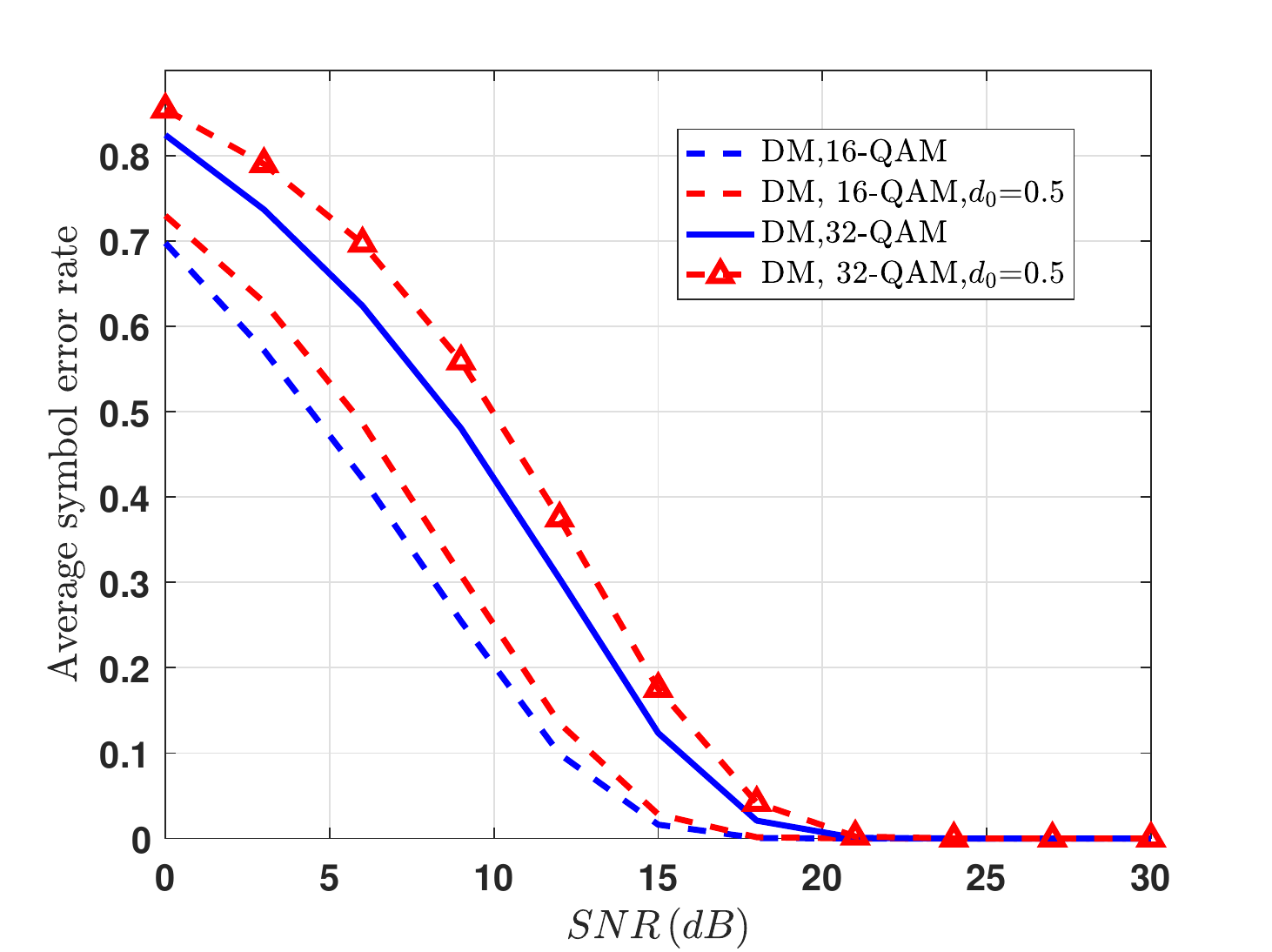}
	\caption{Average SER with respect to the required SNR at the receiver for the proposed $M$-QAM directional modulation precoding when fixed and relaxed detection region designs for $s_{n_4^{}} \in \mathfrak{s_4}$ are considered with $N_t=N_r=10$.}  
	\label{fig:ser:relax}
\end{figure}
\begin{figure}[]
	\centering
	\includegraphics[width=9.5cm]{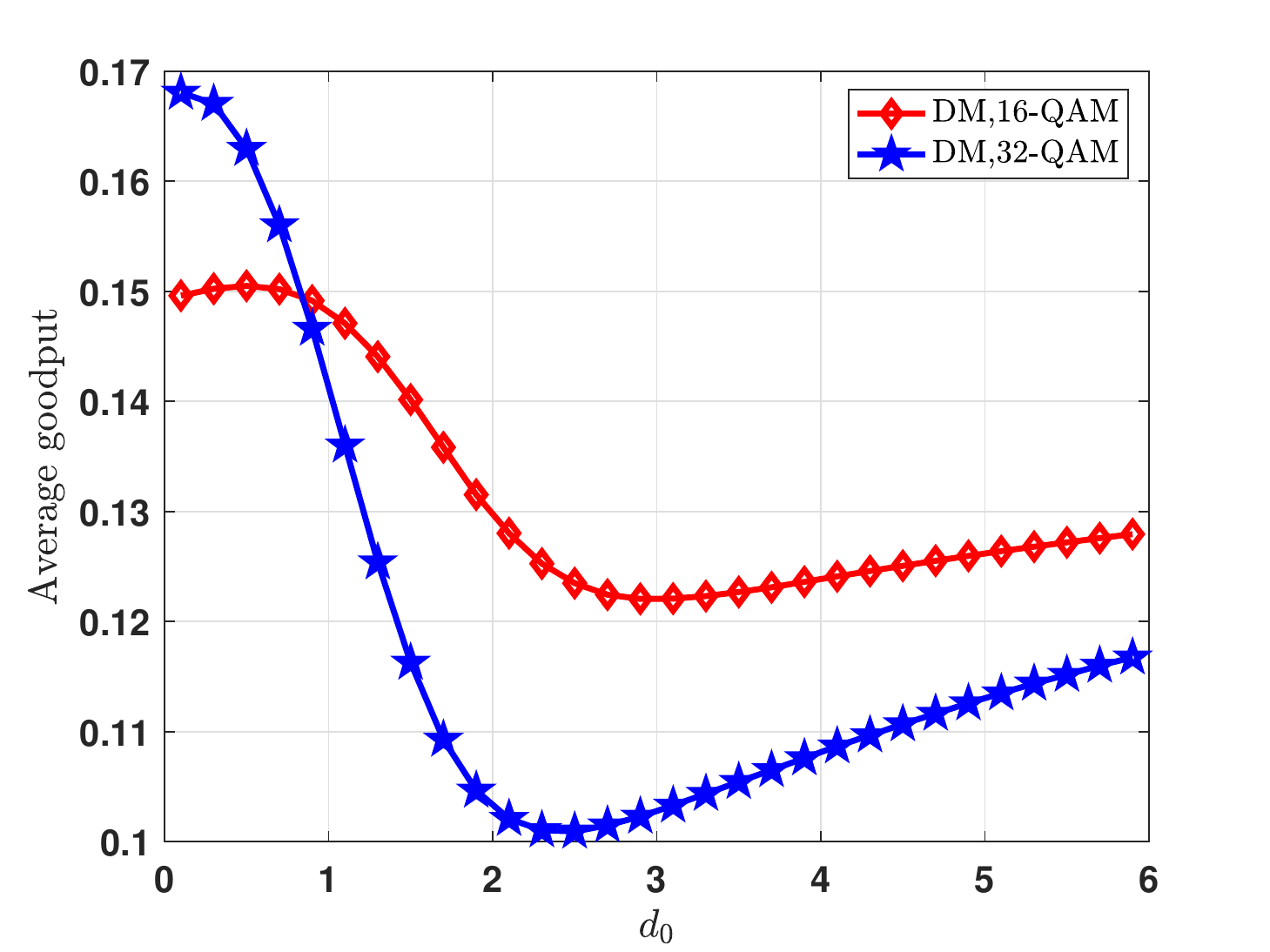}
	\caption{Average goodput over consumed power with respect to $d_0$ for the proposed $16$ and $32$-QAM directional modulation precoding schemes when $N_t=N_r=10$ and $SNR=16$ dB.}  
	\label{fig:gp}
\end{figure}
\begin{figure*}[]
	\centering
	\begin{subfigure}[b]{0.49\textwidth}
		\includegraphics[width=\textwidth]{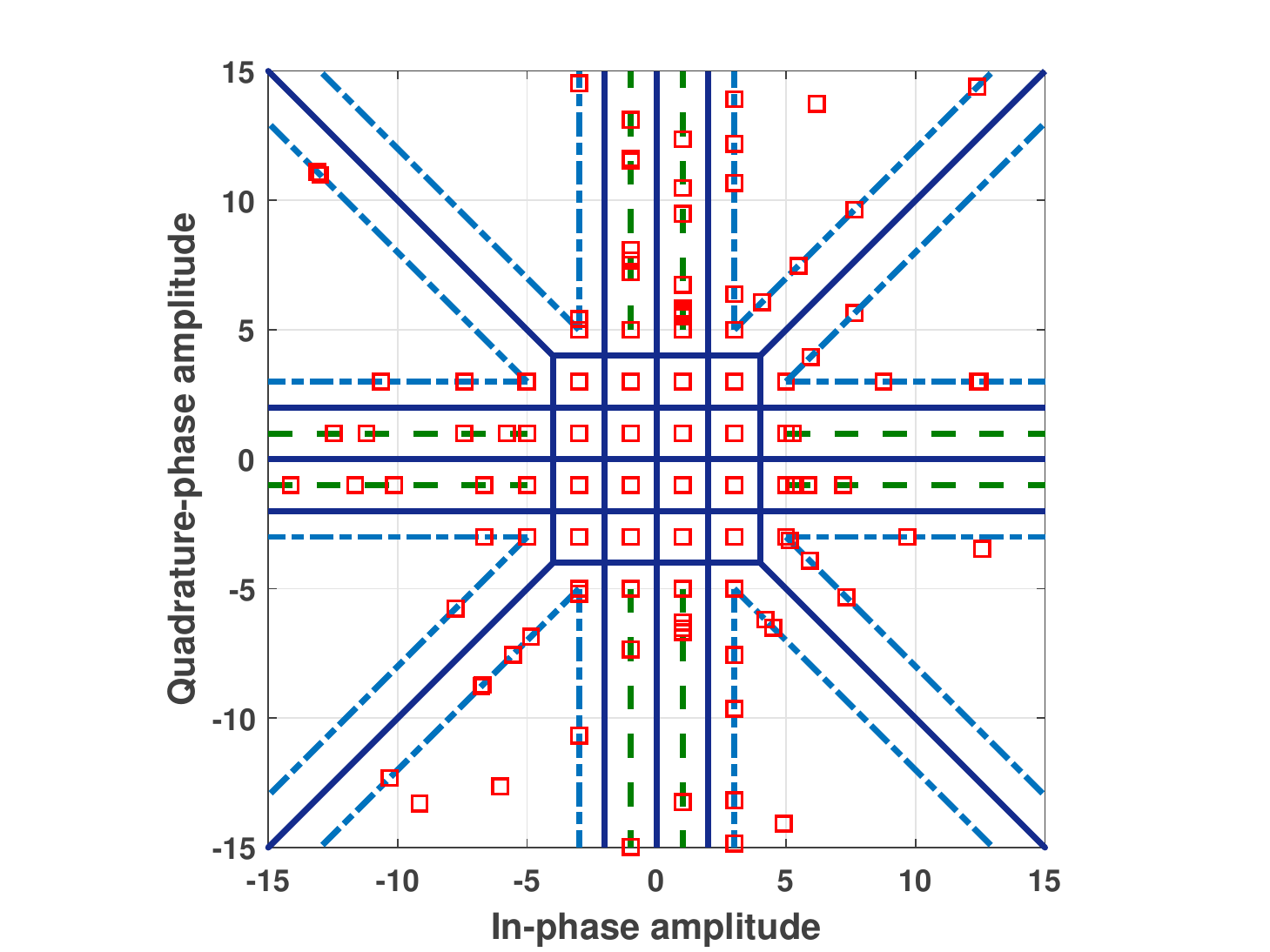}
        \caption{}		
		\label{subfig:32:qam:pre:fixed}
	\end{subfigure}
	\begin{subfigure}[b]{0.49\textwidth}
		\includegraphics[width=\textwidth]{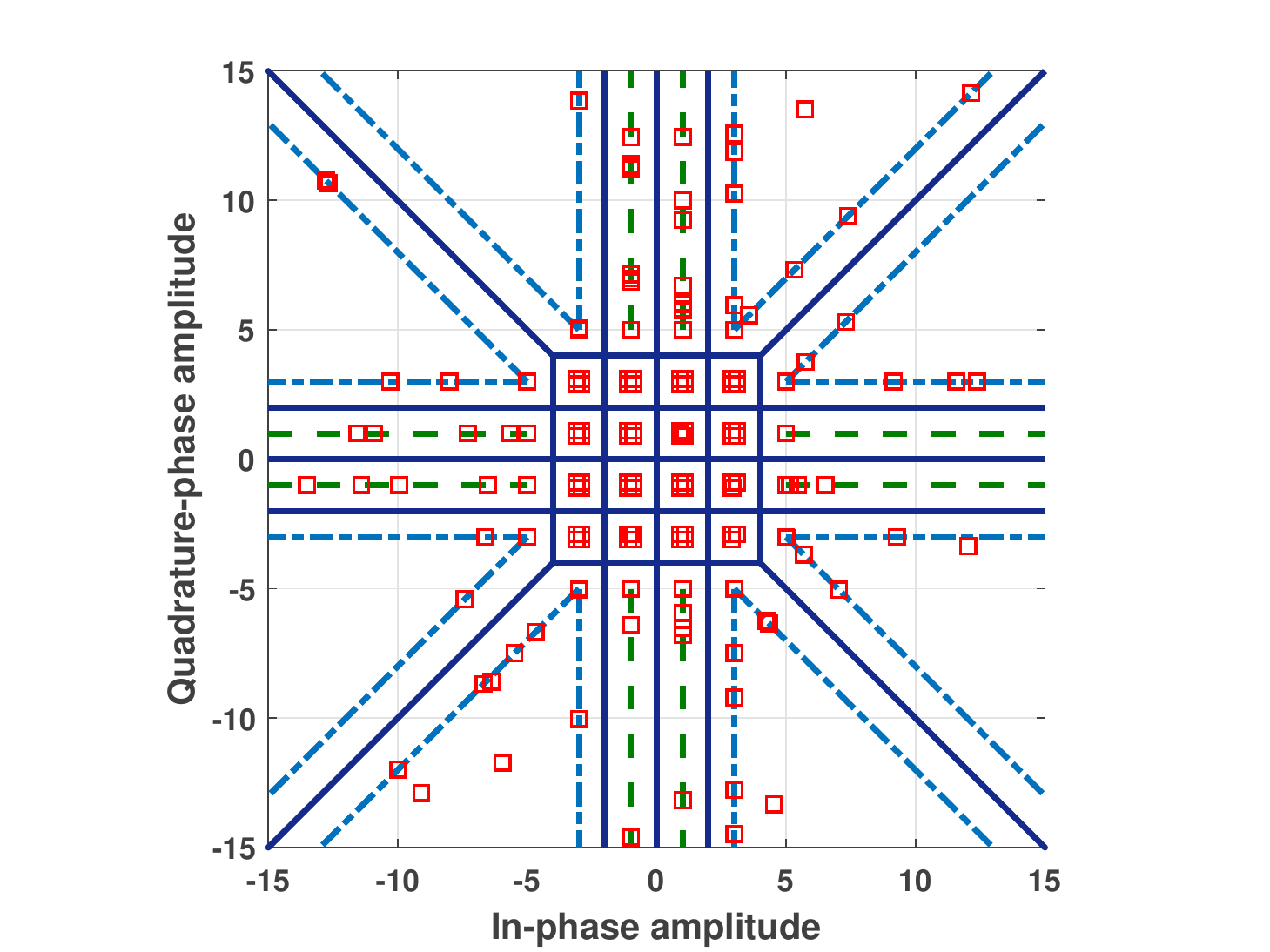}
        \caption{}		
		\label{subfig:32:qam:pre:relaxed}
	\end{subfigure}
	\caption{The constellation of the induced symbols at the receiver for the proposed $32$-QAM directional modulation precoding with $N_t=N_r=500$ and $\gamma =20$ when fixed detection, Fig.~\ref{subfig:32:qam:pre:relaxed}, and relaxed detection with $d_0=0.1$ for $s_{n_4^{}} \in \mathfrak{s_4}$, Fig.~\ref{subfig:32:qam:pre:fixed}, are considered.}
	\label{fig:fixed:relaxed:constellation}
\end{figure*}
In this section, we demonstrate the performance metrics of the proposed methods and compare them with those of the benchmark schemes. We use average over various designed precoders to measure the performance metrics: total power consumption, maximum spatial peak power, SER, and bit error rate (BER). Each designed precoder in the proposed method is used to communicate $N_r$ symbols. In all simulations, channels are considered to be quasi static block Rayleigh fading generated as i.i.d. complex Gaussian random variables with distribution $\mathcal{CN}(0 , 1 )$ and remain fixed during the communication of a group of $N_r$ $M$-QAM symbols. Also, the noise is generated using i.i.d. complex Gaussian random variables with distribution $\mathcal{CN}(0 , {\sigma^2} )$. We assume adaptive coding and modulation in the simulation scenarios and consider specific SNR range in which each modulation order operates.

To save space, we use the acronyms ``DM'', ``Opt LP'', and ``PPM'' in the legend of the figures instead of the terms ``directional modulation'', ``Optimal linear precoding'' method of~\cite{ant:handbook}, and ``spatial peak power minimization'', respectively. In the following, we first mention the benchmark schemes and then proceed with the simulation scenarios.
\subsection{Benchmark Schemes}   \label{sec:bm}
In this part, we mention the zero-forcing~\cite{PPM:Wei} at the transmitter, and optimal linear precoding~\cite{ant:handbook} as the comparison benchmark schemes. 
\subsubsection{ZF}   \label{subsub:zf}
We consider zero-forcing (ZF) at the transmitter~\cite{Lai-U:2004} as one of the benchmark schemes since both directional modulation and ZF use the CSI knowledge at the transmitter to design the precoder and ZF results in interference-free MIMO communication. In the benchmark scheme, we apply the ZF precoder at the transmitter to remove the interference among the transmitted symbol streams. After applying ZF, the received signal at $R$ is
\begin{align}
{{\bf{y}}} = {{\bf{H}}}{\bf{W}}^{} \bf{s}+ {{\bf{n}}}, 
\label{eqn:rec:sig:zf}
\end{align}
where ${\bf{W}}^{} = {\bf{H}}^H{\left( {{\bf{H}}{\bf{H}}^H} \right)^{ - 1}}$ is the precoding vector and vector $\bf{s}$ contains the symbols to be transmitted.
\subsubsection{Optimal linear precoding}   \label{subsub:LP}
The optimal linear precoding design problem using channel state information can be written as~\cite{ant:handbook}
\begin{align}
& \mathop {\min }\limits_{  {\bf{w}}_1,...,{\bf{w}}_{N_r}   } \,\,\, \sum\limits_{i = 1}^{{N_r}} {{{\left\| {{{\bf{w}}_i}} \right\|}^2}} 
\nonumber\\
&\,\, \text{s.t.} \,\, \frac{{{{\left\| {{\bf{h}}_k^T{{\bf{w}}_k}} \right\|}^2}}}{{\sum\limits_{j \ne k}^{{N_r}} {{\bf{h}}_j^T{{\bf{w}}_j}}  + {\sigma ^2}}} \ge {\gamma}, \,\,\, \forall \, k = 1,...,{N_r}.
\label{eqn:LP}
\end{align}
The precoder design problem in~\eqref{eqn:LP} can be solved using semidefinite programming and rank-one relaxation. If the solution to~\eqref{eqn:LP} happens not to be rank-one, randomization can be used to derive a rank-one solution~\cite{relax:MA:2010}. 

The instantaneous spatial peak power minimization version of~\eqref{eqn:LP}, can be cast as
\begin{align}
& \mathop {\min }\limits_{  {\bf{w}}_1,...,{\bf{w}}_{N_r}, {\bf{t}}  } \,\,\, {\left\| {\bf{t}} \right\|^2}
\nonumber\\
&  \,\, \text{s.t.}   \,\,\,\,\,\, {{\bf{w}}_1}{{\bf{E}}_i}{{\bf{w}}_1} + ... + {{\bf{w}}_{{N_r}}}{{\bf{E}}_i}{{\bf{w}}_{{N_r}}} \le {t_i} , \,\,\, \forall \, i = 1,...,{N_t},
\nonumber\\
&   \qquad \,\,\, \frac{{{{\left\| {{\bf{h}}_k^T{{\bf{w}}_k}} \right\|}^2}}}{{\sum\limits_{j \ne k}^{{N_r}} {{\bf{h}}_j^T{{\bf{w}}_j}}  + {\sigma ^2}}} \ge {\gamma} , \,\,\, \forall \, k = 1,...,{N_r},
\label{eqn:LP:PPM}
\end{align}
where ${\bf{t}} = \left[ {{t_1},...,{t_{{N_r}}}} \right]$. A similar approach as in~\eqref{eqn:LP} can be used to solve~\eqref{eqn:LP:PPM}. The spatial peak power is minimized in~\eqref{eqn:LP:PPM} over the precoding vectors ${\bf{w}}_i$, however, in optimal linear precoding, the precoding vectors ${{{\bf{w}}_i}}$ will be multiplied by the $M$-QAM symbols, summed as ${{\bf{y}}_t} = \sum\nolimits_{i = 1}^{{N_r}} {{{\bf{w}}_i}{s_i}}$, and transmitted. As a result, this may change the transmission power of the antennas. To effectively minimize the maximum element of the transmit signal ${{\bf{y}}_t}$, we need to minimize each element of $\overline {\bf{w}}  = \sum\nolimits_{i = 1}^{{N_r}} {{{\bf{w}}_i}}$ since the elements of each ${\bf{w}}_i$ are multiplied by the symbols and then summed up.
\subsection{Simulation Scenarios}   \label{sec:sim_sce}
For the first scenario, we measure the transmitter's average consumed power and spatial peak power for the proposed $M$-QAM directional modulation as well as the benchmark schemes with respect to transmitter's number of antennas, $N_t$. The average total consumed power of the proposed and benchmark schemes versus $N_t$ are shown in Fig.~\ref{fig:pow:ant} for specific system parameters. As we see, the proposed $4$, $8$, $16$, and $32$-QAM directional modulation precoder designs with power minimization consume considerably less power than the ZF and optimal linear precoding schemes for specific range of $N_t$, especially for close values of $N_t$ and $N_r$. As the modulation order decreases, the difference between the consumed power by the proposed and the benchmark schemes increases. For instance when $N_t=N_r=10$ and $SNR=10$ dB, $4$-QAM and $8$-QAM  are respectively $6.07$ and $5.07$ dB below the optimal linear precoding benchmark. Also, $16$-QAM and $32$-QAM are $6.98$ and $4.54$ dB below the benchmark scheme for $SNR=20$ dB. The reason is that in contrast to the conventional ZF and optimal linear precoding, directional modulation takes advantage of the available detection regions of $M$-QAM constellation by designing the precoder based on the symbols. This lets the symbols be placed in the optimal location of the defined regions while satisfying the required SNR at the receiving antennas. 

The MIMO communication systems usually operate in square mode, i.e., equal number of transmit and receive antenna, since the multiplexing gain of the system is the minimum of transmit and receive antennas. Hence, the proposed scheme is a good candidate for MIMO systems since it provides the highest gain compared to optimal linear precoding for close values of $N_t$ and $N_r$, e.g., $N_t=N_r$, as Fig.~\ref{fig:pow:ant} shows. 


We investigate the spatial peak power in the second scenario. The average spatial peak power with respect to $N_t$ is shown in Fig.~\ref{fig:pow:ant:max}. 
The first observation is that as the number of transmit antennas, $N_t$, increases, the transmitter ability to reduce the maximum output power among the power amplifiers increases. The second observation is that for lower modulation orders, the transmitter with spatial peak power minimization is more capable of reducing the level of the power amplifier signal while satisfying the SNR requirements at the receiving antennas. It is seen that the benchmark schemes result in a higher spatial peak power compared to the proposed schemes. 

To analyze the effect of relaxed detection region for $s_{n_4^{}} \in \mathfrak{s_4}$ of $16$-QAM and $32$-QAM constellations introduced in Section~\ref{sec:fre:reg:char}, we have shown the average consumed power of the relaxed design with respect to $N_t$ in Fig.~\ref{fig:pow:ant:relaxed}. As it is illustrated, the relaxed detection region design for $s_{n_4^{}} \in \mathfrak{s_4}$ results in a lower power consumption. Depending on $d_0$, this difference holds for a long range of $N_t$. Interestingly, as $N_t$ increases, the average consumed power of $32$-QAM gets close to that of $16$-QAM in both fixed and relaxed design.

In the next scenario, we measure the average consumed power, average maximum spatial peak power, and the average symbol error rate with respect to the required SNR at the receiver. The average consumed power with respect to the required SNR is shown in Fig.~\ref{subfig:pow_snr} for $N_t=N_r=10$. As it is observed, the consumed power increases consistently with respect to the required SNR. In a wide range of SNRs, the proposed scheme results in a lower power consumption compared to the optimal linear precoding. The average maximum spatial peak power with respect to SNR is shown in Fig.~\ref{subfig:pow_snr_max}. As it is seen, directional modulation results in a lower spatial peak power compared to the optimal linear precoding in a wide range of SNRs. 

To study the effect of relaxed detection region design for $s_{n_4^{}} \in \mathfrak{s_4}$ with respect to SNR, the average total consumed power with respect to SNR is presented in Fig.~\ref{fig:pow:snr:fixed:Nt11} for $N_t=11$ and $N_r=10$. The results shows that the relaxed design results in lower power consumption in low SNR regime. In relatively high SNRs, the consumed powers by the fixed and relaxed designs converge. The convergence SNR depends on the value of $d_0$. This is due to the fact that as the required SNR at the destination increases, the relaxed detection region gets relatively smaller compared to the required SNR and consequently the gain of relaxed design fades out. It is seen that $32$-QAM with relaxed design consumes lower power than $16$-QAM in low SNR since $32$-QAM has four times more constellation points with relaxed design compared to $16$-QAM.   

The average SER with respect to the required SNR is shown in Fig.~\ref{fig:ser}. It is observed that the SER of ZF is close to our scheme since it totally neutralize the interference in the extended detection region design. On the other hand, the optimal linear precoding results in a higher SER compared to the proposed schemes, especially for $M=8,16,32$, since it does not fully mitigate the interference. In addition, the difference in the SER of the proposed method and optimal linear precoding goes higher as the SNR increases. The effect of low density parity check code (LDPC) on BER of the proposed scheme is shown in Fig.~\ref{fig:ber}. 
Next, we investigate the effect of the relaxed detection region on SER. The average SER with respect to SNR is shown in Fig.~\ref{fig:ser:relax}. The relaxed detection region design increases the SER in relatively low SNR regime, however, the SER gets close to the fixed design as SNR increases. 

As was shown, applying the relaxed detection region results in both power reduction and SER increment. To figure out the optimal value for relaxation, $d_0$, we need to consider a metric which captures both power consumption and SER. To do so, let us consider the goodput over the total consumed power defined in~\eqref{eqn:gp}, as the performance metric. The value of $\eta$ with respect to $d_0$ is presented in Fig.~\ref{fig:gp}. It is observed that there exists a value of $d_0$ for both $16$ and $32$-QAM where it is possible to setup an optimal trade-off between power consumption and SER.  

The benefit of the proposed relaxed detection region becomes more clear as we consider Figures~\ref{fig:pow:ant:relaxed} and~\ref{fig:ser:relax} together. In this example, we see that for $N_t=N_r=10$ at $SNR=20$~dB, the relaxed design results in power reduction for $32$-QAM design while the SER is almost the same as the fixed design. Hence, for specific values of $d_0$ and modulation type, the relaxed design can provide power gain for negligible SER increment.

At the end, we present an example of the communicated $32$-QAM constellation points at the receiver for both relaxed and fixed detection region designs of $s_n \in \mathfrak{s_4}$ in Figures~\ref{subfig:32:qam:pre:fixed} and~\ref{subfig:32:qam:pre:relaxed}. As we see, many symbols go above the required signal level at the destination, which can results in lower SER. Particularly, we see that the optimally precoded points in the sets $\mathfrak{s_5}$ and $\mathfrak{s_6}$ are concentrated at the boundary of the extended detection regions. 
\subsection{Soft Decisions}   \label{subsec:soft}
In the numerical example of Fig.~\ref{fig:ber} with LDPC coding, the demodulator takes hard decisions and feeds these decisions to the channel decoder. However, nothing prevents, in principle, the computation of soft decisions. Significantly, in order for the LDPC code to perform well, such soft decisions must be computed based on an accurate model for the conditional probability density of the received signal point given each hypothetically induced symbol (or bit) at the receiving antennas. This probability density in turn is non-trivial to find when directional modulation is used, because in addition to the channel noise (which is modeled as additive Gaussian here), the optimization problem in~\eqref{eqn:pre:gen4} itself introduces an uncertainty into the received sample, even in the noise-free case. 

More specifically, consider the $4$-QAM case and suppose the symbol $(-1,-1)$ is destined for the $i$-th terminal with $\gamma=1$. The received point, say $x = {\bf{h}}_i^T{\bf{w}}$, at this terminal may then (neglecting the noise) lie anywhere in the quadrant ${\mathop{\rm Re}\nolimits} \left( x \right) \le 1$, ${\mathop{\rm Im}\nolimits} \left( x \right) \le 1$; and depending on the symbols destined for the other terminals, $x$ may be more or less far away from $(-1,-1)$. The problem is now that first, the application of a decoding metric tailored to additive Gaussian noise may yield seriously over-pessimistic log-likelihood ratios for this symbol (should $x$ be far away from $(-1,-1)$); second, that the probability density of $x$ depends on symbols aimed at the other terminals and hence is unknown.  

Given this observation, we have to leave the design of proper soft-decision algorithms for directional modulation to future work. 
\section{Conclusions}   \label{sec:con}
In this paper, we proposed the concept of extended and relaxed detection regions in directional modulation to design the optimal directional modulation precoders for $M$-QAM constellations with total and spatial peak power minimization. We transformed the problems into convex ones and proposed a path-following interior point algorithm to solve them. Through directional modulation precoding, symbols are placed in the optimal location of extended and relaxed detection region. Results showed that the suggested $M$-QAM directional modulation precoding consumes less power than the conventional ZF in all ranges of SNR, and less power than the optimal linear precoding for relatively middle and high SNR regimes. We demonstrated that this difference in power consumption increases as the modulation order decreases. Also, the results showed that directional modulation results in lower spatial peak power in all SNR ranges compared to ZF as well as lower spatial peak power compared to optimal linear precoding in a wide range of SNRs. The results showed that the difference between the average spatial peak power of the proposed and the benchmark schemes increases as the modulation order decreases. 

In addition, we saw that the interference-free communication capability of directional modulation, considering fixed design for inner constellation points of $M=16,32$, results in lower SER at the receiver compared to optimal linear precoding. This difference in SER increases for $M=8,16,32$ as SNR increases. Considering that the proposed scheme can provide both lower power consumption and SER depending on SNR and modulation order, it can provide communication at a specific rate using a lower amount of power. In case of adaptive coding and modulation, it is possible to switch to a higher order modulation, due to lower SER, using a lower power consumption at the transmitter. 

We demonstrated that precoder design using the relaxed detection region for inner constellation points of $16$-QAM and $32$-QAM results in lower power consumption and SER increment in a specific range of SNR, which depends on the relaxation value. We showed that depending on relaxation value and modulation type, the relaxed detection region design results in lower power consumption and SER with negligible increment compared to fixed design in relatively high SNR. It was shown than the goodput, considering both power consumption and SER, over the total power consumption can be optimized for a specific relaxation value.   
\appendices
\section{Transformation of $32$-QAM \\ Optimal Precoder Design Problem}   \label{app:trans}
To simplify~\eqref{eqn:dm:32-QAM1:pow:min}, first, we stack the constraints of~\eqref{eqn:dm:32-QAM1:pow:min} as
\begin{subequations}
	\begin{align}
	& \mathop {\min }\limits_{{\bf{w}}} \,\,\, {\left\| {\bf{w}} \right\|^2}
	\nonumber\\
	&  \,\, \text{s.t.}   \,\,\,\,\,\,  {\rm{Re}}\left( {{{\bf{H}}_2}{\bf{w}}} \right) = {{\bf{s}}_{re_2^{}}}, \, {\rm{Im}}\left( {{{\bf{S}}_{{_2}}}} \right){\rm{Im}}\left( {{{\bf{H}}_2}{\bf{w}}} \right) \ge {\bf{s}}_{{im_2^{}}}^{\circ 2},
	\label{subeq:pre:des:24}
	\\
	&             \qquad \,\,\, {\rm{Re}}\left( {{{\bf{S}}_3}} \right){\rm{Re}}\left( {{{\bf{H}}_3}{\bf{w}}} \right) \ge {\bf{s}}_{{re_3^{}}}^{\circ 2}, \,  {\rm{Im}}\left( {{{\bf{H}}_3}{\bf{w}}} \right) = {\bf{s}}_{{im_3^{}}},
	\label{subeq:pre:des:22}
	\\
	&             \qquad \,\,\,  {\rm{Re}}\left( {{{\bf{H}}_4}{\bf{w}}} \right) = {{\bf{s}}_{r{e_4}}}, \, {\rm{Im}}\left( {{{\bf{H}}_4}{\bf{w}}} \right) = {{\bf{s}}_{i{m_4}}},
	\label{subeq:pre:des:29}
	\\
	&             \qquad \,\,\,  3 \times {{\bf{1}}_5} \le {\rm{Im}}\left( {{{\widetilde {\bf{H}}}_5}{\bf{w}}} \right) \le {\rm{Re}}\left( {{{\widetilde {\bf{H}}}_5}{\bf{w}}} \right) - 2 \times {{\bf{1}}_5},
	\label{subeq:pre:des:25}
	\\
	&             \qquad \,\,\,  {\rm{Re}}\left( {{{\widetilde {\bf{H}}}_6}{\bf{w}}} \right) + 2 \times {{\bf{1}}_6} \le {\rm{Im}}\left( {{{\widetilde {\bf{H}}}_6}{\bf{w}}} \right),
	\label{subeq:pre:des:26}
	\\
	&             \qquad \,\,\, {\rm{Re}}\left( {{{\widetilde {\bf{H}}}_6}{\bf{w}}} \right) \ge 3 \times {{\bf{1}}_6},
	\label{subeq:pre:des:27}				
	\end{align}
	\label{eqn:dm:32-QAM2:pow:min}%
\end{subequations}
where ${{\bf{s}}_{re_j^{}}}$ and ${{\bf{s}}_{im_j^{}}}$ are the vectors that respectively stack the real and imaginary parts of the symbols $s_{n_j^{}} \in \mathfrak{s_{j}}$ multiplied by $\sqrt \gamma$, ${{\bf{1}}_j}$ is a $Card(\mathfrak{s_{j}}) \times 1$ vector with elements multiplied by $\sqrt \gamma$, and ${\bf{S}}_j$ is a diagonal matrix with diagonal entries as $s_{n_j^{}} \in \mathfrak{s_{j}}$. Next, we proceed to remove the real and imaginary operators from~\eqref{eqn:dm:32-QAM2:pow:min}. Similar as in~\cite{DM:JSTSP:kalantari:2016}, we have 
\begin{align}
{\rm{Re}}\left( {{{\bf{H}}_j}{\bf{w}}} \right) = {{\bf{H}}_{ja}}\widetilde {\bf{w}}, \,\, {\rm{Im}}\left( {{{\bf{H}}_j}{\bf{w}}} \right) = {{\bf{H}}_{jb}}\widetilde {\bf{w}},
\label{eqn:rel:ima:Hw}
\end{align}
where $\widetilde {\bf{w}} = {\left[ {{\rm{Re}}\left( {\bf{w}}^T \right),{\rm{Im}}\left( {\bf{w}}^T \right)} \right]^T}$, ${{\bf{H}}_{ja}} = \left[ {{\rm{Re}}\left( {{{\bf{H}}_j}} \right), - {\rm{Im}}\left( {{{\bf{H}}_j}} \right)} \right]$, ${{\bf{H}}_{jb}} = \left[ {{\rm{Im}}\left( {{{\bf{H}}_j}} \right),{\rm{Re}}\left( {{{\bf{H}}_j}} \right)} \right]$, and ${\left\| {\widetilde {\bf{w}}} \right\|^2} = {\left\| {\bf{w}} \right\|^2}$. Using the results in~\eqref{eqn:rel:ima:Hw}, we can reformulate~\eqref{eqn:dm:32-QAM2:pow:min} as   
\begin{subequations}
	\begin{align}
	& \mathop {\min }\limits_{{\bf{w}}} \,\,\, {\left\| {\bf{w}} \right\|^2}
	\nonumber\\
	&  \,\, \text{s.t.}   \,\,\,\,\,\, {{\bf{H}}_{2a}}{\bf{w}} = {{\bf{s}}_{r{e_2^{}}}}, \, {\rm{Im}}\left( {{{\bf{S}}_2}} \right){{\bf{H}}_{2b}}{\bf{w}} \ge {\bf{s}}_{i{m_2^{}}}^{\circ 2},
	\label{subeq:pre:des:34}
	\\
	&             \qquad \,\,\, {\rm{Re}}\left( {{{\bf{S}}_3}} \right){{\bf{H}}_{3a}}{\bf{w}} \ge {\bf{s}}_{r{e_3^{}}}^{\circ 2}, \, {{\bf{H}}_{3b}}{\bf{w}} = {{\bf{s}}_{i{m_3^{}}}},
	\label{subeq:pre:des:32}
	\\
	&             \qquad \,\,\, {{\bf{H}}_{4a}}{\bf{w}} = {{\bf{s}}_{r{e_4^{}}}}, \,  {{\bf{H}}_{4b}}{\bf{w}} = {{\bf{s}}_{i{m_4^{}}}},
	\label{subeq:pre:des:39}
	\\
	&             \qquad \,\,\,  3 \times {{\bf{1}}_5} \le {\widetilde {\bf{H}}_{5b}}{\bf{w}} \le {\widetilde {\bf{H}}_{5a}}{\bf{w}} - 2 \times {{\bf{1}}_5},
	\label{subeq:pre:des:35}
	\\
	&             \qquad \,\,\, {\widetilde {\bf{H}}_{6a}}{\bf{w}} + 2 \times {{\bf{1}}_6} \le {\widetilde {\bf{H}}_{6b}}{\bf{w}}, \, {\widetilde {\bf{H}}_{6a}}{\bf{w}} \ge 3 \times {{\bf{1}}_6},
	\label{subeq:pre:des:37}				
	\end{align}
	\label{eqn:pre:gen3:pow:min}%
\end{subequations}
Stacking the constraints of~\eqref{eqn:pre:gen3:pow:min} yields
\begin{align}
& \mathop {\min }\limits_{{\bf{w}}} \,\,\, {\left\| {\bf{w}} \right\|^2} \, \text{s.t.}  \,\, {\bf{Aw}} \ge {\bf{a}}, \, {\bf{Bw}} = {\bf{b}},
\label{eqn:pre:32:qam:fix:final1}
\end{align}
where {
\begin{align}
&{\bf{A}} = \left( \begin{array}{l}
{\rm{Im}}\left( {{{\bf{S}}_2}} \right){{\bf{H}}_{2b}}\\
{\rm{Re}}\left( {{{\bf{S}}_3}} \right){{\bf{H}}_{3a}}\\
{\widetilde {\bf{H}}_{5a}} - {\widetilde {\bf{H}}_{5b}}\\
{\widetilde {\bf{H}}_{5b}}\\
{\widetilde {\bf{H}}_{6b}} - {\widetilde {\bf{H}}_{6a}}\\
{\widetilde {\bf{H}}_{6a}}
\end{array} \right),\,\,{\bf{a}} = \left( \begin{array}{l}
{\bf{s}}_{i{m_2^{}}}^{\circ 2}\\
{\bf{s}}_{r{e_3^{}}}^{\circ 2}\\
2\times{{\bf{1}}_5}\\
3\times{{\bf{1}}_5}\\
2\times{{\bf{1}}_6}\\
3\times{{\bf{1}}_6}
\end{array} \right),
\nonumber\\
&{\bf{B}} = \left( \begin{array}{l}
{{\bf{H}}_{2a}}\\
{{\bf{H}}_{3b}}\\
{{\bf{H}}_{4a}}\\
{{\bf{H}}_{4b}}
\end{array} \right),\,\, {\bf{b}} = \left( \begin{array}{l}
{{\bf{s}}_{r{e_2^{}}}}\\
{{\bf{s}}_{i{m_3^{}}}}\\
{{\bf{s}}_{r{e_4^{}}}}\\
{{\bf{s}}_{i{m_4^{}}}}
\end{array} \right).
\label{eqn:def}
\end{align}}
If we consider the relaxed detection region for the points $s_{n_4^{}} \in \mathfrak{s_4}$, we again reach to a convex problem similar as~\eqref{eqn:pre:32:qam:fix:final} where $\bf{A}$, $\bf{a}$, $\bf{B}$ and $\bf{b}$ are as follows:
{
\begin{align}
&{\bf{A}} = \left( \begin{array}{l}
{\rm{Im}}\left( {{{\bf{S}}_2}} \right){{\bf{H}}_{2b}}\\
{\rm{Re}}\left( {{{\bf{S}}_3}} \right){{\bf{H}}_{3a}}\\
{{\bf{H}}_{4a}}\\
{{\bf{H}}_{4b}}\\
- {{\bf{H}}_{4a}}\\
- {{\bf{H}}_{4b}}\\
{\widetilde {\bf{H}}_{5a}} - {\widetilde {\bf{H}}_{5b}}\\
{\widetilde {\bf{H}}_{5b}}\\
{\widetilde {\bf{H}}_{6b}} - {\widetilde {\bf{H}}_{6a}}\\
{\widetilde {\bf{H}}_{6a}}
\end{array} \right),\,\,  {\bf{a}} = \left( \begin{array}{l}
{\bf{s}}_{i{m_2}}^{ \circ 2}\\
{\bf{s}}_{r{e_3}}^{ \circ 2}\\
{{\bf{s}}_{r{e_4}}} - {{\bf{d}}_0}\\
{{\bf{s}}_{i{m_4}}} - {{\bf{d}}_0}\\
- {{\bf{s}}_{r{e_4}}} - {{\bf{d}}_0}\\
- {{\bf{s}}_{i{m_4}}} - {{\bf{d}}_0}\\
2 \times {{\bf{1}}_5}\\
3 \times {{\bf{1}}_5}\\
2 \times {{\bf{1}}_6}\\
3 \times {{\bf{1}}_6}
\end{array} \right), 
\nonumber\\
&{\bf{B}} = \left( \begin{array}{l}
{{\bf{H}}_{2a}}\\
{{\bf{H}}_{3b}}
\end{array} \right),\,\, {\bf{b}} = \left( \begin{array}{l}
{{\bf{s}}_{r{e_2}}}\\
{{\bf{s}}_{i{m_3}}}
\end{array} \right).
\label{eqn:def:rel}
\end{align}}
\bibliographystyle{IEEEtran}

\begin{thebibliography}{10}
	\providecommand{\url}[1]{#1}
	\csname url@samestyle\endcsname
	\providecommand{\newblock}{\relax}
	\providecommand{\bibinfo}[2]{#2}
	\providecommand{\BIBentrySTDinterwordspacing}{\spaceskip=0pt\relax}
	\providecommand{\BIBentryALTinterwordstretchfactor}{4}
	\providecommand{\BIBentryALTinterwordspacing}{\spaceskip=\fontdimen2\font plus
		\BIBentryALTinterwordstretchfactor\fontdimen3\font minus
		\fontdimen4\font\relax}
	\providecommand{\BIBforeignlanguage}[2]{{%
			\expandafter\ifx\csname l@#1\endcsname\relax
			\typeout{** WARNING: IEEEtran.bst: No hyphenation pattern has been}%
			\typeout{** loaded for the language `#1'. Using the pattern for}%
			\typeout{** the default language instead.}%
			\else
			\language=\csname l@#1\endcsname
			\fi
			#2}}
	\providecommand{\BIBdecl}{\relax}
	\BIBdecl
	
	\bibitem{cisco}
	``Cisco visual networking index: Forecast and methodology, 2015-2020,'' Cisco
	Systems, Tech. Rep., 2015.
	
	\bibitem{OFDMA:1971}
	S.~Weinstein and P.~Ebert, ``Data transmission by frequency-division
	multiplexing using the discrete fourier transform,'' \emph{IEEE Trans.
		Commun. Technol.}, vol.~19, no.~5, pp. 628--634, Oct. 1971.
	
	\bibitem{TDMA:1995}
	D.~D. Falconer, F.~Adachi, and B.~Gudmundson, ``Time division multiple access
	methods for wireless personal communications,'' \emph{IEEE Commun. Mag.},
	vol.~33, no.~1, pp. 50--57, Jan. 1995.
	
	\bibitem{Spencer:2004}
	Q.~Spencer, A.~Swindlehurst, and M.~Haardt, ``Zero-forcing methods for downlink
	spatial multiplexing in multiuser {MIMO} channels,'' \emph{IEEE Trans. Signal
		Process.}, vol.~52, no.~2, pp. 461--471, Feb. 2004.
	
	\bibitem{Sidiropoulos:2006}
	N.~Sidiropoulos, T.~Davidson, and Z.-Q. Luo, ``Transmit beamforming for
	physical-layer multicasting,'' \emph{IEEE Trans. Signal Process.}, vol.~54,
	no.~6, pp. 2239--2251, Jun. 2006.
	
	\bibitem{Babakhani:2008}
	A.~Babakhani, D.~Rutledge, and A.~Hajimiri, ``Transmitter architectures based
	on near-field direct antenna modulation,'' \emph{IEEE J. Solid-State
		Circuits}, vol.~43, no.~12, pp. 2674--2692, Dec. 2008.
	
	\bibitem{Daly:2009}
	M.~Daly and J.~Bernhard, ``Directional modulation technique for phased
	arrays,'' \emph{IEEE Trans. Antennas Propag.}, vol.~57, no.~9, pp.
	2633--2640, Sep. 2009.
	
	\bibitem{Daly:2010}
	M.~Daly, E.~Daly, and J.~Bernhard, ``Demonstration of directional modulation
	using a phased array,'' \emph{IEEE Trans. Antennas Propag.}, vol.~58, no.~5,
	pp. 1545--1550, May 2010.
	
	\bibitem{DM:JSTSP:kalantari:2016}
	A.~Kalantari, M.~Soltanalian, S.~Maleki, S.~Chatzinotas, and B.~Ottersten,
	``Directional modulation via symbol-level precoding: A way to enhance
	security,'' \emph{IEEE J. Sel. Topics Signal Process.}, vol.~10, no.~8, pp.
	1478--1493, Dec. 2016.
	
	\bibitem{Kalantari:psk:relax}
	A.~Kalantari, C.~Tsinos, M.~Soltanalian, S.~Chatzinotas, W.-K. Ma, and
	B.~Ottersten, ``Low peak power {MIMO} directional modulation transmitter
	design for relaxed phase {M-PSK} modulation,'' in \emph{European signal
		Processing Conference}, Kos, Greece, Aug. 2017.
	
	\bibitem{Kalantari:MQAM}
	A.~Kalantari, C.~Tsinos, M.~Soltanalian, S.~Chatzinotas, and B.~Ottersten,
	``{MIMO} directional modulation {M-QAM} precoding for transceivers
	performance enhancement,'' in \emph{IEEE International workshop on Signal
		Processing advances in Wireless Communications (SPAWC)}, Sapporo, Japan,
	2017.
	
	\bibitem{lowcom:pre}
	J.~Krivochiza, A.~Kalantari, S.~Chatzinotas, and B.~Ottersten, ``Low complexity
	symbol-level design for linear precoding systems,'' in \emph{Symposium on
		Inf. Theory and Signal Process. in the Benelux}, Delft, Netherlands, May
	2017.
	
	\bibitem{Masouros:2009}
	C.~Masouros and E.~Alsusa, ``Dynamic linear precoding for the exploitation of
	known interference in {MIMO} broadcast systems,'' \emph{IEEE Trans. Wireless
		Commun.}, vol.~8, no.~3, pp. 1396--1404, Mar. 2009.
	
	\bibitem{CDMA:Mas:2010}
	------, ``Soft linear precoding for the downlink of {DS/CDMA} communication
	systems,'' \emph{IEEE Trans. Veh. Technol.}, vol.~59, no.~1, pp. 203--215,
	Jan. 2010.
	
	\bibitem{cons:2015}
	M.~Alodeh, S.~Chatzinotas, and B.~Ottersten, ``Constructive multiuser
	interference in symbol level precoding for the {MISO} downlink channel,''
	\emph{IEEE Trans. Signal Process.}, vol.~63, no.~9, pp. 2239--2252, May 2015.
	
	\bibitem{con:glob:2015}
	------, ``Constructive interference through symbol level precoding for
	multi-level modulation,'' in \emph{IEEE Global Commun. Conf. (GLOBECOM)}, CA,
	San Diego, Dec. 2015.
	
	\bibitem{multi:adaptive:TWC}
	------, ``Symbol-level multiuser {MISO} precoding for multi-level adaptive
	modulation,'' \emph{IEEE Trans. Wireless Commun.}, vol.~PP, no.~99, pp. 1--1,
	2017.
	
	\bibitem{pre:survey}
	\BIBentryALTinterwordspacing
	M.~Alodeh, D.~Spano, A.~Kalantari, C.~Tsinos, D.~Christopoulos, S.~Chatzinotas,
	and B.~Ottersten, ``Symbol-level and multicast precoding for multiuser
	multiantenna downlink: A state-of-the-art, classification and challenges,''
	\emph{Commun. Surveys Tuts.} [Online]. Available:
	\url{https://arxiv.org/pdf/1601.02788.pdf}
	\BIBentrySTDinterwordspacing
	
	\bibitem{mobile:carbon:2011}
	A.~Fehske, G.~Fettweis, J.~Malmodin, and G.~Biczok, ``The global footprint of
	mobile communications: The ecological and economic perspective,'' \emph{IEEE
		Commun. Mag.}, vol.~49, no.~8, pp. 55--62, Aug. 2011.
	
	\bibitem{mob:ene:2010}
	L.~M. Correia, D.~Zeller, O.~Blume, D.~Ferling, Y.~Jading, I.~Godor, G.~Auer,
	and L.~V.~D. Perre, ``Challenges and enabling technologies for energy aware
	mobile radio networks,'' \emph{IEEE Commun. Mag.}, vol.~48, no.~11, pp.
	66--72, Nov. 2010.
	
	\bibitem{Masouros:2015}
	C.~Masouros and G.~Zheng, ``Exploiting known interference as green signal power
	for downlink beamforming optimization,'' \emph{IEEE Trans. Signal Process.},
	vol.~63, no.~14, pp. 3628--3640, Jul. 2015.
	
	\bibitem{EE:const:2016}
	M.~Alodeh, S.~Chatzinotas, and B.~Ottersten, ``Energy-efficient symbol-level
	precoding in multiuser {MISO} based on relaxed detection region,'' \emph{IEEE
		Trans. Wireless Commun.}, vol.~15, no.~5, pp. 3755--3767, May 2016.
	
	\bibitem{amplifier:1981}
	A.~A.~M. Saleh, ``Frequency-independent and frequency-dependent nonlinear
	models of {TWT} amplifiers,'' \emph{IEEE Trans. Wireless Commun.}, vol.~29,
	no.~11, pp. 1715--1720, Nov. 1981.
	
	\bibitem{Erik:2012:CE}
	S.~K. Mohammed and E.~G. Larsson, ``Single-user beamforming in large-scale
	{MISO} systems with per-antenna constant-envelope constraints: The doughnut
	channel,'' \emph{IEEE Trans. Wireless Commun.}, vol.~11, no.~11, pp.
	3992--4005, Nov. 2012.
	
	\bibitem{CE:Ma:2014}
	J.~Pan and W.~K. Ma, ``Constant envelope precoding for single-user large-scale
	{MISO} channels: Efficient precoding and optimal designs,'' \emph{IEEE J.
		Sel. Topics Signal Process.}, vol.~8, no.~5, pp. 982--995, Oct. 2014.
	
	\bibitem{danilo:glob:2016}
	D.~Spano, M.~Alodeh, S.~Chatzinotas, and B.~Ottersten, ``Per-antenna power
	minimization in symbol-level precoding,'' in \emph{IEEE Global Commun. Conf.
		(GLOBECOM)}, Washington, DC, USA, Dec. 2016.
	
	\bibitem{slp:nonlin}
	------, ``Symbol-level precoding for the nonlinear multiuser {MISO} downlink
	channel,'' \emph{IEEE Trans. Signal Process.}, vol.~66, no.~5, pp.
	1331--1345, Mar. 2018.
	
	\bibitem{lin:pre:TWC:2012}
	Y.~Wu, M.~Wang, C.~Xiao, Z.~Ding, and X.~Gao, ``Linear precoding for {MIMO}
	broadcast channels with finite-alphabet constraints,'' \emph{IEEE Trans.
		Wireless Commun.}, vol.~11, no.~8, pp. 2906--2920, Aug. 2012.
	
	\bibitem{LTE}
	{The 3rd Generation Partnership Project (3GPP)}. {LTE}.
	http://www.3gpp.org/technologies/keywords-acronyms/98-lte.
	
	\bibitem{Nesterov1994}
	Y.~Nesterov and A.~Nemirovskii, \emph{Interior-point Polynomial Algorithms in
		Convex Programming}, ser. Studies in Applied Mathematics.\hskip 1em plus
	0.5em minus 0.4em\relax Philadelphia, PA: Society for Industrial and Applied
	Mathematics (SIAM), 1994.
	
	\bibitem{Dennis1996NMf}
	J.~E. Dennis, Jr. and R.~B. Schnabel, \emph{Numerical Methods for Unconstrained
		Optimization and Nonlinear Equations}, ser. Classics in Applied
	Mathematics.\hskip 1em plus 0.5em minus 0.4em\relax Philadelphia, PA: SIAM,
	1996, vol.~16.
	
	\bibitem{ant:handbook}
	M.~Bengtsson and B.~Ottersten, \emph{Handbook of Antennas in Wireless
		Communications}.\hskip 1em plus 0.5em minus 0.4em\relax CRC Press, 2001, ch.
	Optimal and suboptimal transmit beamforming.
	
	\bibitem{PPM:Wei}
	W.~Yu and T.~Lan, ``Transmitter optimization for the multi-antenna downlink
	with per-antenna power constraints,'' \emph{IEEE Trans. Signal Process.},
	vol.~55, no.~6, pp. 2646--2660, Jun. 2007.
	
	\bibitem{Lai-U:2004}
	L.-U. Choi and R.~Murch, ``A transmit preprocessing technique for multiuser
	{MIMO} systems using a decomposition approach,'' \emph{IEEE Trans. Wireless
		Commun.}, vol.~3, no.~1, pp. 20--24, Jan. 2004.
	
	\bibitem{relax:MA:2010}
	Z.-Q. Luo, W.-K. Ma, A.~M.-C. So, Y.~Ye, and S.~Zhang, ``Semidefinite
	relaxation of quadratic optimization problems,'' \emph{IEEE Signal Process.
		Mag.}, vol.~27, no.~3, pp. 20--34, May 2010.
	
\end{thebibliography}

\end{document}